\documentclass[aps,superscriptaddress,prl,showpacs,twocolumn]{revtex4-2}

\usepackage[colorlinks = true,
            linkcolor = blue,
            urlcolor  = blue,
            citecolor = blue,
            anchorcolor = blue]{hyperref}
\usepackage{amsmath,amssymb}
\usepackage{commath}
\usepackage{bm,bbm}
\usepackage{physics}
\usepackage{tikz}
 \usepackage{booktabs}
\usepackage{graphicx}
\usepackage{xcolor}
\usepackage{epstopdf}
\usepackage{CJKutf8}
\usepackage{orcidlink}
\usepackage[normalem]{ulem}

\newcommand{\jl}[1]{{\color{black} {#1}}}

\begin{document}

\title{\jl{Spin-Orbit-Mediated Magnetization Transfer in the Fermi-Hubbard Model}}

\author{Jialiang Tang$^{\orcidlink{0009-0001-6420-6492}}$}
\affiliation{Quantum Advanced Research Center (QuARC), CSIC, 28049, Madrid, Spain}
\affiliation{Instituto de Ciencia de Materiales de Madrid (ICMM), CSIC, 28049, Madrid, Spain}
\affiliation{\mbox{Departamento de Física Teórica de la Materia Condensada, Universidad Autónoma de Madrid, 28049 Madrid, Spain}}

\author{Ruoqian Xu$^{\orcidlink{0009-0005-3242-9674}}$}
\affiliation{Quantum Advanced Research Center (QuARC), CSIC, 28049, Madrid, Spain}
\affiliation{Instituto de Ciencia de Materiales de Madrid (ICMM), CSIC, 28049, Madrid, Spain}
\affiliation{\mbox{Departamento de Física Teórica de la Materia Condensada, Universidad Autónoma de Madrid, 28049 Madrid, Spain}}

\author{Gloria Platero$^{\orcidlink{0000-0001-8610-0675}}$}
\affiliation{Quantum Advanced Research Center (QuARC), CSIC, 28049, Madrid, Spain}
\affiliation{Instituto de Ciencia de Materiales de Madrid (ICMM), CSIC, 28049, Madrid, Spain}

\author{Yue Ban$^{\orcidlink{0000-0003-1764-4470}}$}
\email{yue.ban@csic.es}
\affiliation{Quantum Advanced Research Center (QuARC), CSIC, 28049, Madrid, Spain}
\affiliation{Instituto de Ciencia de Materiales de Madrid (ICMM), CSIC, 28049, Madrid, Spain}

\author{Xi Chen$^{\orcidlink{0000-0003-4221-4288}}$}
\email{xi.chen@csic.es}
\affiliation{Quantum Advanced Research Center (QuARC), CSIC, 28049, Madrid, Spain}
\affiliation{Instituto de Ciencia de Materiales de Madrid (ICMM), CSIC, 28049, Madrid, Spain}
\date{\today}

\begin{abstract}
We demonstrate spin--orbit-coupling (SOC) catalysis of magnetization-sector transfer in the half-filled Fermi--Hubbard model. Without SOC, conservation of total magnetization confines the dynamics to the initial sector, preventing an antiferromagnetically correlated state from reaching the polarized ferromagnetic sector under a Zeeman ramp. A transient SOC control lifts this obstruction by coupling otherwise disconnected magnetization sectors and converting symmetry-protected crossings into avoided crossings, thereby opening a finite-gap pathway for magnetic-state conversion. We further accelerate the SOC-enabled dynamics beyond the adiabatic timescale using gradient-ascent pulse engineering (GRAPE) and a variational quantum circuit. Both approaches achieve rapid magnetization transfer in one- and  two-dimensional Fermi--Hubbard lattices. \jl{Our results establish transient SOC as a controllable catalyst for
overcoming symmetry-imposed dynamical constraints, while optimal control
sets the timescale for the resulting magnetization-sector transfer.}
\end{abstract}

\maketitle

\textit{Introduction.--}
The Fermi--Hubbard (FH) model provides a paradigmatic setting for
strongly correlated fermions, capturing the interplay of particle
motion, interactions, and magnetic order
~\cite{imada1998metal,esslinger2010fermi,arovas2022hubbard}.
At half filling with repulsive interactions, antiferromagnetic
correlations emerge through superexchange and have been observed in
cold-atom quantum simulators
~\cite{cheuk2016observation,mazurenko2017cold}.
Adiabatic evolution offers a natural route for preparing and
transforming such correlated states
~\cite{farhi2000quantumcomputationadiabaticevolution,
albash2018adiabatic,simon2011quantum,
kim2024adiabaticstatepreparationquantum}.
A qualitatively different challenge arises, however, when an exact
symmetry partitions the Hilbert space into dynamically disconnected
sectors. In this case, slowing down the evolution cannot remove the
obstruction: if all available Hamiltonian generators preserve the
symmetry, no pulse shaping can connect states belonging to different
sectors~\cite{Zhuang2020SymmetryProtected}. 

 \begin{figure}[t]
 \centering
\includegraphics[width=\linewidth]{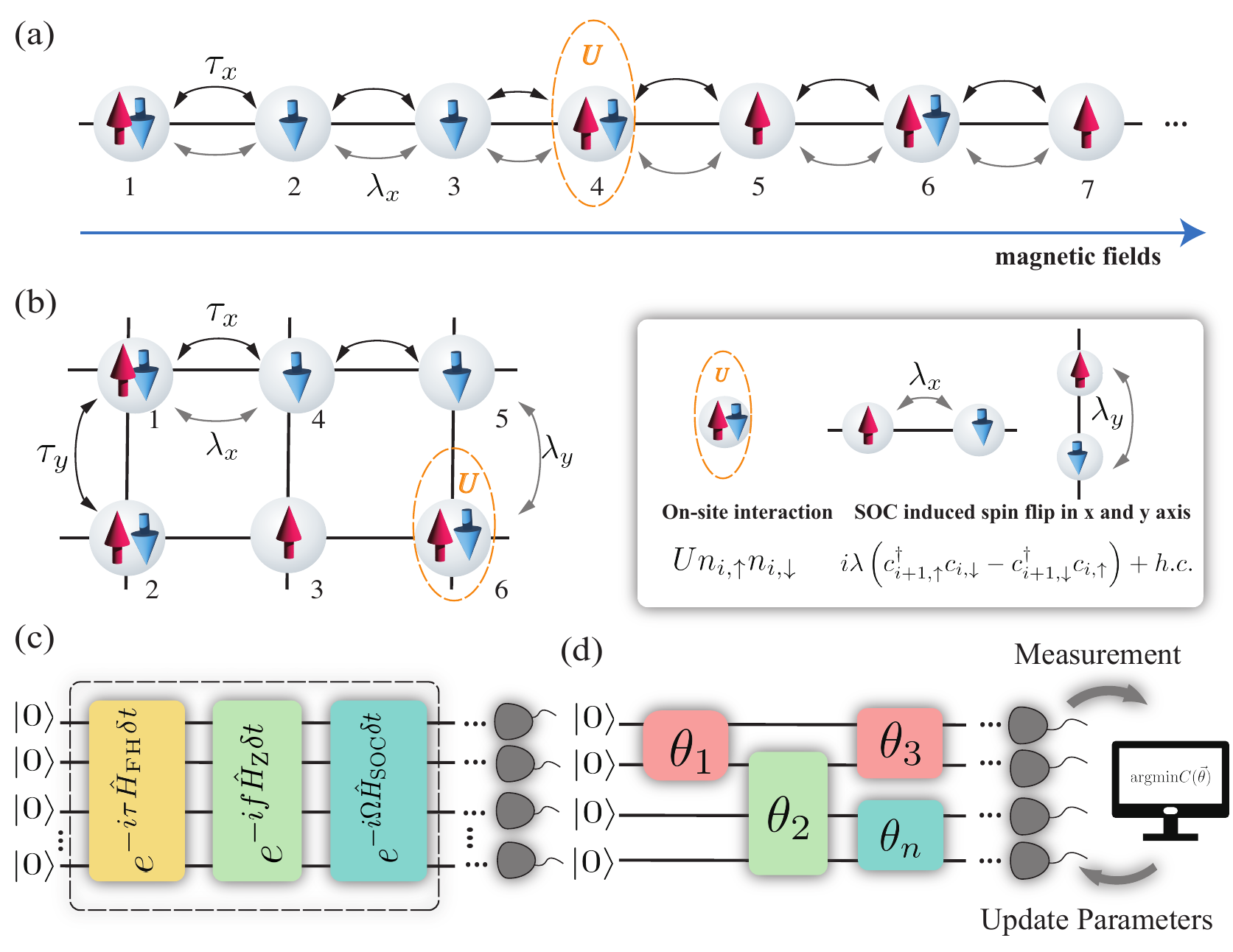}
\caption{Fermi--Hubbard models and control architecture.
(a) One-dimensional chain with hopping $\tau_x$, on-site interaction $U$,
longitudinal Zeeman driving, and SOC-induced spin-flip hopping $\lambda_x$.
(b) Two-dimensional lattice with hopping amplitudes $\tau_{x,y}$ and SOC
strengths $\lambda_{x,y}$ along the two lattice directions. The inset highlights the on-site interaction and SOC-assisted spin-flip process.
(c) Trotterized digital representation of the SOC-assisted dynamics in terms
of hopping, interaction, Zeeman, and SOC evolution blocks.
(d) Quantum--classical variational control loop, where measured observables
are used to update the layer-dependent circuit parameters $\theta_i$.}
\label{fig: model}
 \end{figure}

This situation arises naturally in magnetization control of the
half-filled FH model. The hopping, interaction, and longitudinal
Zeeman terms all conserve the total magnetization, so that an
antiferromagnetically correlated state in the zero-magnetization sector
cannot evolve into a fully polarized sector even when the
instantaneous ground state changes its magnetization along a Zeeman
ramp. Spin--orbit coupling (SOC) provides a physical route around this
constraint. Synthetic SOC and spin-dependent tunneling have been
engineered in a variety of atomic and condensed-matter platforms
~\cite{PhysRevLett.111.066801,PhysRevB.88.165136,
galitski2013spin,PhysRevLett.123.160404,
smidman2017superconductivity,borisenko2016direct}.
Here, SOC-induced spin-flip hopping breaks magnetization conservation,
couples otherwise disconnected sectors, and converts
symmetry-protected crossings into avoided crossings. When applied only
transiently, SOC therefore acts as a catalytic control that opens a
finite-gap pathway without modifying the initial or final Hamiltonian,
in analogy with catalyst Hamiltonians in quantum annealing
~\cite{PhysRevA.103.022608,PhysRevA.110.042609,
ghosh2024exponentialspeedupquantumannealing}.

Opening such a pathway does not by itself make the transfer fast.
We therefore combine SOC with quantum optimal control. At the
Hamiltonian level, we use gradient-ascent pulse engineering (GRAPE)
~\cite{Khaneja2005GRAPE} to optimize the Zeeman and SOC controls.
At the circuit level, we map the same dynamics to a Trotterized
quantum circuit after fermion-to-qubit transformation
~\cite{Lloyd1996,Barends_2015,Barends_2016} and optimize its
layer-dependent pulse areas through variational circuit learning
~\cite{PRXQuantum.2.010101,TangyouPRR}.
Recent progress in programmable FH simulators, including
ultracold-atom and semiconductor quantum-dot platforms
~\cite{mazurenko2017cold,3nx4-bnyy,Hensgens_2017,
hartnett2026fastaccuratehighresolutionsimulation,
nigmatullin2025experimental}
makes both control settings experimentally relevant.
Thus, SOC determines which magnetic sectors are dynamically accessible,
whereas optimal control determines how rapidly they can be connected.

In this Letter, we propose that transient SOC converts the
symmetry-protected crossings into avoided crossings, thereby enabling
magnetization-sector transfer in the half-filled FH model [Fig.~\ref{fig: model}]. Hamiltonian-level GRAPE and direct circuit optimization then strongly
accelerate this SOC-enabled pathway in both 1D and 2D geometries.

\textit{Fermi--Hubbard model and SOC-assisted sector transfer.--}
We consider the spinful FH model at half filling
~\cite{imada1998metal,esslinger2010fermi}, as illustrated in
Fig.~\ref{fig: model}(a),
\begin{equation}
\hat H_{\rm FH}
=
-\tau\sum_{\langle i,j\rangle,\sigma}
\left(
c_{i,\sigma}^{\dagger}c_{j,\sigma}
+\mathrm{h.c.}
\right)
+
U\sum_i n_{i,\uparrow}n_{i,\downarrow}.
\label{eq:hopping}
\end{equation}
where $c_{i,\sigma}^{\dagger}$ ($c_{i,\sigma}$) creates (annihilates) an electron with spin $\sigma$ on site $i$, $\tau$ is the nearest-neighbor hopping amplitude, $U$ is the on-site Coulomb repulsion, and $n_{i,\sigma}=c_{i,\sigma}^{\dagger} c_{i,\sigma}$ is the local number operator. At half filling and for $U\gg\tau$, virtual hopping generates
antiferromagnetic superexchange with characteristic scale
$J_{\rm ex}\simeq4\tau^2/U$~\cite{mazurenko2017cold}.
The spin-conserving FH Hamiltonian preserves the total
$z$-magnetization,
\begin{equation}
\hat{\mathcal M}
\equiv
\hat S_{\rm tot}^{z}
=
\frac{1}{2}\sum_i
\left(
n_{i,\uparrow}-n_{i,\downarrow}
\right),
\label{eq:mag_symmetry}
\end{equation}
which satisfies
$[\hat H_{\rm FH},\hat{\mathcal M}]=0$.
This observable can be reconstructed from spin-resolved site
occupations~\cite{mazurenko2017cold,Cheuk2015,Cheuk2016}.
Accordingly, the Hilbert space decomposes as
$\mathcal H=\bigoplus_{\mathcal M}\mathcal H_{\mathcal M}$,
where $\mathcal H_{\mathcal M}$ denotes the subspace with fixed
magnetization $\mathcal M$.  The dynamics is initialized in the antiferromagnetically correlated
FH ground state, prepared by adiabatically interpolating from a
staggered N\'eel Hamiltonian to the interacting FH Hamiltonian;
details are given in ~\cite{SM}.

To drive the system toward spin polarization, we consider
\begin{equation}
\hat H(t)
=
\hat H_{\rm FH}
+
f(t)\hat H_Z
+
\lambda(t)\hat H_{\rm SOC},
\label{eq:total}
\end{equation}
where
$\hat H_Z=
B\sum_i
\left(
n_{i,\uparrow}-n_{i,\downarrow}
\right)
=
2B\hat{\mathcal M}$ and  $\hat H_{\rm SOC}
=
i\lambda_{\rm SOC}\sum_{\langle i,j\rangle}
\left(
c_{j,\uparrow}^{\dagger}c_{i,\downarrow}
-
c_{j,\downarrow}^{\dagger}c_{i,\uparrow}
\right)
+\mathrm{h.c.}$. Here, \(\lambda_{\rm SOC}\) sets the SOC energy scale.
In gate-defined quantum-dot arrays, the hopping and SOC terms correspond
respectively to spin-conserving and spin-flip tunneling amplitudes, which
can be controlled electrically~\cite{fernandez2024flying}.
The ordinary hopping is spin conserving and therefore preserves the
magnetization sector, whereas the SOC-induced spin-flip tunneling
changes the spin projection and couples neighboring magnetization
sectors.
The two controls obey
\begin{equation}
[\hat H_Z,\hat{\mathcal M}]=0,
\qquad
[\hat H_{\rm SOC},\hat{\mathcal M}]\neq0,
\label{eq:commutators}
\end{equation}
and therefore play qualitatively different roles.
The Zeeman field shifts the relative energies of different
magnetization sectors without coupling them, so that
\(\hat H_{\rm FH}+f(t)\hat H_Z\) leaves the dynamics confined to the
initial sector. By contrast, SOC-induced spin-flip hopping couples
otherwise disconnected sectors and removes this symmetry-based
dynamical obstruction.

\begin{figure}[t]
\centering
\includegraphics[width=\linewidth]{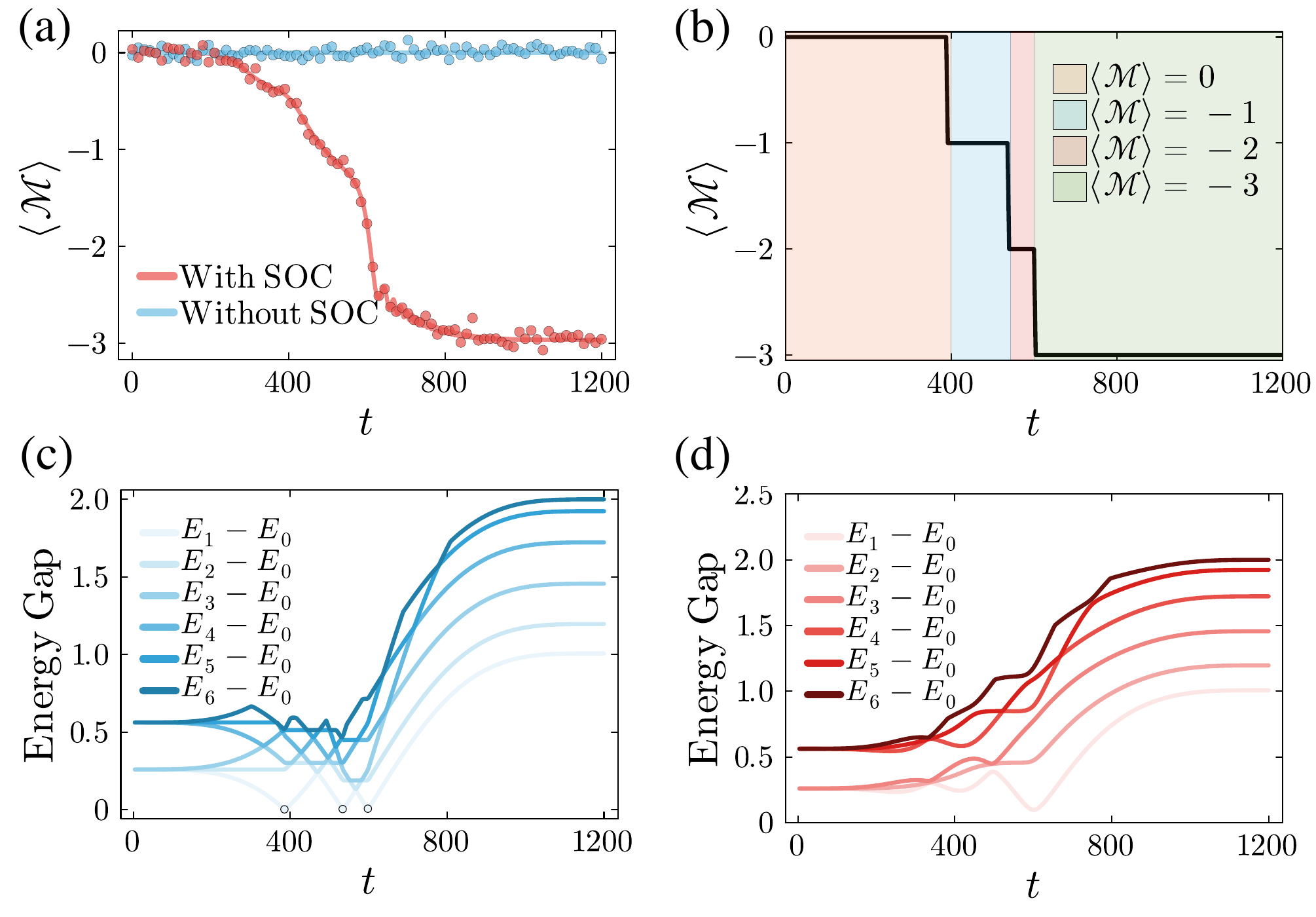}
\caption{(a) Magnetization dynamics along the reference ramp with and without
SOC. Solid lines denote Hamiltonian evolution, while symbols show the
corresponding digitized dynamics. (b) Magnetization quantum number $\mathcal M$ for the instantaneous ground state for $\lambda(t) = 0$, obtained by exact diagonalization.  (c,d) \jl{Instantaneous low-energy excitation spectrum,
\(E_n(t)-E_0(t)\) for \(n=1,\ldots,6\), along the adiabatic path
without and with SOC, respectively.}   
Parameters: $U=7$, $\lambda_{\rm SOC}=1$, and $t_f=1200$ for the adiabatic reference.} 
\label{fig:energy_gap}
\end{figure} 

We choose a smooth Zeeman ramp $f(t)
=
\sin^2\![
\frac{\pi}{2}
\sin^2\!(
\frac{\pi t}{2t_f}
)]$, with $f(0)=0$ and $f(t_f)=1$~\cite{Narendraprapplied21}, where $t_f$ is the total evolution time.
The transient SOC control is taken as a normalized pulse envelope inherited from the Zeeman ramp, \(\lambda(t)=\dot f(t)/\max_t|\dot f(t)|\). This choice ensures that the SOC vanishes at both endpoints while
remaining active during the ramp.
Such spin-flip hopping can be engineered in synthetic-SOC optical-lattice
platforms \cite{Galitski2013SpinOrbitQuantumGases,PhysRevLett.109.085303,PhysRevA.98.023623}, with close counterparts in condensed-matter systems exhibiting Rashba-type SOC~\cite{smidman2017superconductivity,borisenko2016direct}.
Gate-defined quantum-dot arrays provide a complementary solid-state
platform for FH dynamics and electrically controlled spin-flip
tunneling~\cite{Hensgens_2017,fernandez2024flying}.

The sector-transfer mechanism is illustrated in
Fig.~\ref{fig:energy_gap} for a half-filled six-site FH chain.
Without SOC, the state remains confined to the initial
$\mathcal M=0$ sector, with
$\langle\hat{\mathcal M}\rangle=0$ throughout the ramp
[Fig.~\ref{fig:energy_gap}(a)], even though the instantaneous ground
state changes successively as
\begin{equation}
\mathcal M
=
0\rightarrow-1\rightarrow-2\rightarrow-3.
\end{equation}
Because these sectors are uncoupled for $\lambda(t)=0$, the changes
occur through exact level crossings and gap closings
[Figs.~\ref{fig:energy_gap}(b,c)].
A transient SOC pulse instead hybridizes neighboring sectors,
converts the crossings into avoided crossings, and opens finite gaps
[Fig.~\ref{fig:energy_gap}(d)].

Near a crossing between the $\mathcal M$ and $\mathcal M-1$ sectors,
the relevant SOC matrix element is
$g_{\mathcal M}
=
\langle\psi_{\mathcal M-1}|
\hat H_{\rm SOC}
|\psi_{\mathcal M}\rangle $,
where $|\psi_{\mathcal M}\rangle$ and
$|\psi_{\mathcal M-1}\rangle$ denote the corresponding bare eigenstates. In the weak-SOC regime,
a two-level projection then gives, 
$ \Delta_{\mathcal M}(t_c)
\simeq
2\tilde{\lambda}_{\rm SOC}(t_c)|g_{\mathcal M}|$, where $\tilde{\lambda}_{\rm SOC}(t_c)=\lambda(t_c)\lambda_{\rm SOC}$.
This linear scaling is verified by exact diagonalization at weak SOC
~\cite{SM}. Near each avoided crossing, the detuning between the two relevant
magnetization sectors is approximately linear in time,
$\delta_{\mathcal M}(t)
\simeq
v_{\mathcal M}(t-t_c)$ with
$v_{\mathcal M}
=
|
\frac{d}{dt}
[
E_{\mathcal M}^{(0)}(t)
-
E_{\mathcal M-1}^{(0)}(t)
]|_{t=t_c}$,
where $E_{\mathcal M}^{(0)}$ denotes the corresponding diabatic sector energy in the absence of SOC, and $t_c$ is bare-sector crossing position.
The local adiabaticity is therefore characterized by
$\Delta^2_{\mathcal M}/v_{\mathcal M}$.
Among the successive avoided crossings, the
$\mathcal M=-2\rightarrow-3$ transition has the smallest value,
identifying it as the dominant dynamical bottleneck \cite{SM}.
This picture is closely related to recent Landau--Zener analyses of
finite FH systems~\cite{triadu2025probing}, although the
origin of the gap is different: here, SOC itself creates the avoided
crossing by coupling otherwise disconnected magnetization sectors.

At larger SOC strengths, multilevel hybridization produces
nonperturbative corrections to the weak-coupling expression, while the
avoided crossings remain open and continue to provide a connected
finite-gap pathway between magnetization sectors.
Because SOC vanishes at both endpoints and acts only by opening this
otherwise forbidden pathway, it serves as a catalyst for
magnetization-sector transfer.

The same symmetry-based mechanism extends naturally to two dimensions.
For the lattice in Fig.~\ref{fig: model}(b), we take
\begin{equation}
\hat H^{2D}(t)
=
\hat H_{\rm FH}^{2D}
+
f(t)\hat H_Z
+
\sum_{\alpha=x,y}
\lambda_{\alpha}(t)\hat H_{\rm SOC}^{\alpha},
\label{eq:2d_total}
\end{equation}
where
$\hat H_{\rm FH}^{2D}=
-\sum_{\alpha=x,y}
\tau_{\alpha}
\sum_{\langle i,j\rangle_{\alpha},\sigma}
\left(
c_{i,\sigma}^{\dagger}c_{j,\sigma}
+
\mathrm{h.c.}
\right)
+
U\sum_i n_{i,\uparrow}n_{i,\downarrow}$ and $\hat H_{\rm SOC}^{\alpha}
=
i \lambda_{\rm SOC} \sum_{\langle i,j\rangle_{\alpha}}
\left(
c_{j,\uparrow}^{\dagger}c_{i,\downarrow}
-
c_{j,\downarrow}^{\dagger}c_{i,\uparrow}
\right)
+
\mathrm{h.c.}$ contain the corresponding hopping and
spin-flip terms along the two lattice directions, with amplitudes
\(\tau_{\alpha}\) and \(\lambda_{\alpha}\) ($\alpha=x,y$). Importantly, the symmetry structure is unchanged:
the FH and Zeeman terms preserve $\hat{\mathcal M}$, whereas the SOC
terms couple different magnetization sectors.  The relevant energy-gap structure and SOC-induced avoided crossing are shown in~\cite{SM}.

\begin{figure}[t]
\centering
\includegraphics[width=\linewidth]{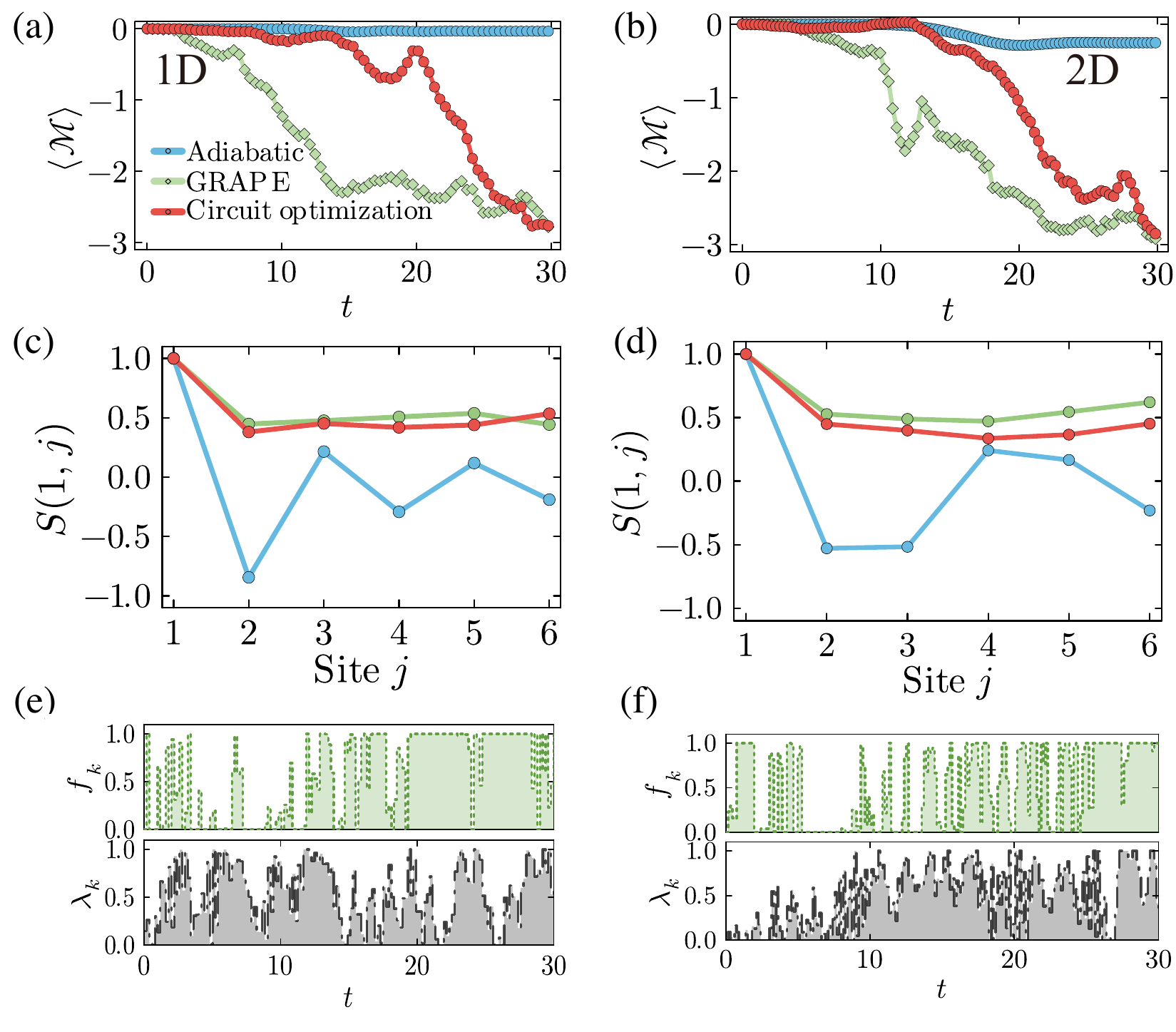}
\caption{
Optimization of SOC-assisted magnetization-sector transfer in
1D (left column) and 2D (right column) FH systems.
(a),(b) Magnetization dynamics for the adiabatic evolution,
GRAPE optimization, and optimized digital circuit.
(c),(d) Final connected spin correlations, showing the conversion from
antiferromagnetic to predominantly ferromagnetic correlations.
(e),(f) GRAPE-optimized Zeeman and SOC amplitudes $f_k$ and $\lambda_k$,
with the FH parameters $\tau$ and $U$ held fixed.
For the 1D chain, $N_T=150$, $\delta t=0.2$, and $t_f=30$;
for the $2\times3$ lattice, $N_T=200$, $\delta t=0.15$,
and $t_f=30$.
}
\label{fig:optimal_control}
\end{figure}

\textit{Hamiltonian-level optimal control.--}
Although SOC removes the symmetry obstruction by opening a finite-gap
pathway between magnetization sectors, adiabatic following can remain
slow because of the small SOC-induced gaps. We therefore use
gradient-ascent pulse engineering (GRAPE)~\cite{Khaneja2005GRAPE} to
optimize the Zeeman and SOC controls while keeping the FH parameters
$\tau$ and $U$ fixed.

We divide the evolution time \(t_f\) into \(N_T\) intervals of
duration \(\delta t=t_f/N_T\), with
\begin{equation}
\hat H_k
=
\tau\hat H_{\rm h}
+
U \hat H_{\rm C}
+
f_k\hat H_Z
+
\lambda_k\hat H_{\rm SOC},
\label{eq:grape_hamiltonian}
\end{equation}
where $\hat H_{\rm h}$ and $\hat H_{\rm C}$ denote the hopping and
interaction terms in Eq.~\eqref{eq:hopping}. The control parameters
$\boldsymbol{\theta}=\{f_k,\lambda_k\}_{k=1}^{N_T}$ are optimized by
minimizing 
\begin{equation}
\mathcal L(\boldsymbol{\theta})
=
\langle\psi(t_f)|\hat{\mathcal M}|\psi(t_f)\rangle ,
\qquad
|\psi(t_f)\rangle
=
\hat U(t_f)|\psi_0\rangle ,
\label{eq:grape_objective}
\end{equation}
with
\(
\hat U(t_f)
=
\prod_{k=N_T}^{1}
\exp[-i\hat H_k\delta t]
\).
For the negatively polarized target, minimizing $\mathcal L$ drives the
system toward the fully polarized sector. Numerical details are given in
the Supplemental Material~\cite{SM}. 
Fig.~\ref{fig:optimal_control} compares the reference adiabatic
evolution with the optimized dynamics. In the 1D chain, an evolution
time $t_f=30$ is too short for adiabatic following, whereas GRAPE drives
the system close to the fully polarized sector within the same time
[Fig.~\ref{fig:optimal_control}(a)]. The same behavior persists for the
$2\times3$ lattice [Fig.~\ref{fig:optimal_control}(b)], demonstrating
that the acceleration is not specific to the one-dimensional spectrum.

To quantify the transfer, we define
\begin{equation}
\mu(t)
=
\frac{
|\langle\hat{\mathcal M}(t)\rangle|
}{
|\mathcal M_{\rm tar}|
},
\label{eq:magnetization_accuracy}
\end{equation}
\jl{where \(\mathcal M_{\rm tar}=-3\) denotes the target magnetization for the six-site systems.}
At a common threshold $\mu\geq0.967$, the 1D protocol \jl{based in GRAPE} reaches the
target at $t_f=30$, compared with $t_f=1200$ for the adiabatic reference,
corresponding to a \jl{forty-fold} reduction in evolution time.
For the $2\times3$ lattice, the corresponding time is reduced from $t_f=200$ for the adiabatic reference to $t_f=30$ under GRAPE
optimization, yielding a \(6.7\)-fold reduction.


The final connected spin--spin correlations, $S(i,j)=\langle S_i^z S_j^z\rangle-\langle S_i^z\rangle\langle S_j^z\rangle$ become predominantly positive, as shown in Figs.~\ref{fig:optimal_control}(c,d),  confirming the emergence of
ferromagnetic correlations in the optimized final state.  \jl{Figs.~\ref{fig:optimal_control}(e,f) show the optimized Zeeman and SOC pulses, with the Zeeman
and SOC amplitudes constrained to the same ranges as in the reference protocol. The resulting piecewise-constant control profiles are
compatible with standard GRAPE implementations based on time-discretized
control amplitudes~\cite{Khaneja2005GRAPE,Machnes2011}. In practice, the time discretization can be chosen according to the
bandwidth and response time of the available Zeeman- and SOC-control
channels.} 
Further acceleration can be obtained by additionally optimizing the
hopping and interaction strengths or by allowing larger control
amplitudes~\cite{SM}; such enlarged control spaces are motivated by the increasing
programmability of optical-lattice FH simulators
~\cite{3nx4-bnyy}.

\textit{Variational quantum-circuit optimization.--}
To assess the SOC-assisted mechanism in a digital setting, we next
formulate the protocol directly at the circuit level
~\cite{Lloyd1996,alam2025programmabledigitalquantumsimulation,tang2024exploring}.
After the Jordan--Wigner transformation, the controlled Hamiltonian in
Eq.~\eqref{eq:grape_hamiltonian} is mapped onto
\begin{equation}
\hat H_Q(t)
=
\sum_m g_m(t)\hat H_m ,
\end{equation}
where \(\hat H_m\) are the qubit representations of the hopping,
interaction, Zeeman, and SOC terms, with corresponding amplitudes
\(g_m(t)\). 
The circuit starts from the same correlated half-filled FH ground state
used in the Hamiltonian-level simulations, prepared by a fixed-depth
CD-inspired variational circuit~\cite{SM}.
The subsequent evolution is approximated
by a first-order Trotter product,
\begin{equation}
\hat U_Q(t_f)
\simeq
\prod_{k=1}^{N_T}
\prod_m
\exp\!\left[
-i g_m(t_k)\hat H_m\delta t
\right],
\label{eq:trotter_general}
\end{equation}
with $t_f=N_T\delta t$.  The agreement between the digitized and Hamiltonian evolutions in
Fig.~\ref{fig:energy_gap}(a) shows that the chosen Trotterization
captures the relevant sector-transfer dynamics.

We then promote the layer-dependent pulse areas to variational
parameters,
\begin{equation}
\hat U_{\rm var}(\boldsymbol{\theta})
=
\prod_{k=1}^{N_T}
\prod_m
\exp\!\left[
-i\theta_{k,m}\hat H_m
\right],
\label{eq:variational_trotter_circuit}
\end{equation}
and optimize them in the variational circuits~\cite{PRXQuantum.2.010101,TangyouPRR}, using the same final-magnetization objective. Unlike GRAPE, which optimizes the control amplitudes before
digitization, the variational circuit optimizes the corresponding pulse
areas directly in the Trotterized representation.
The learned circuit reproduces the accelerated transfer in both
geometries [Figs.~\ref{fig:optimal_control}(a,b)] and yields
predominantly ferromagnetic final correlations
[Figs.~\ref{fig:optimal_control}(c,d)].
At the final time, it reaches $\mu=0.959$ in 1D and $\mu=0.967$ for the
$2\times3$ lattice, comparable to the Hamiltonian-level optimization.
These results show that the SOC-enabled pathway remains effective after
fermion-to-qubit mapping, digitization, and direct circuit-level
optimization.

\begin{figure}[t]
\centering
\includegraphics[width=\linewidth]{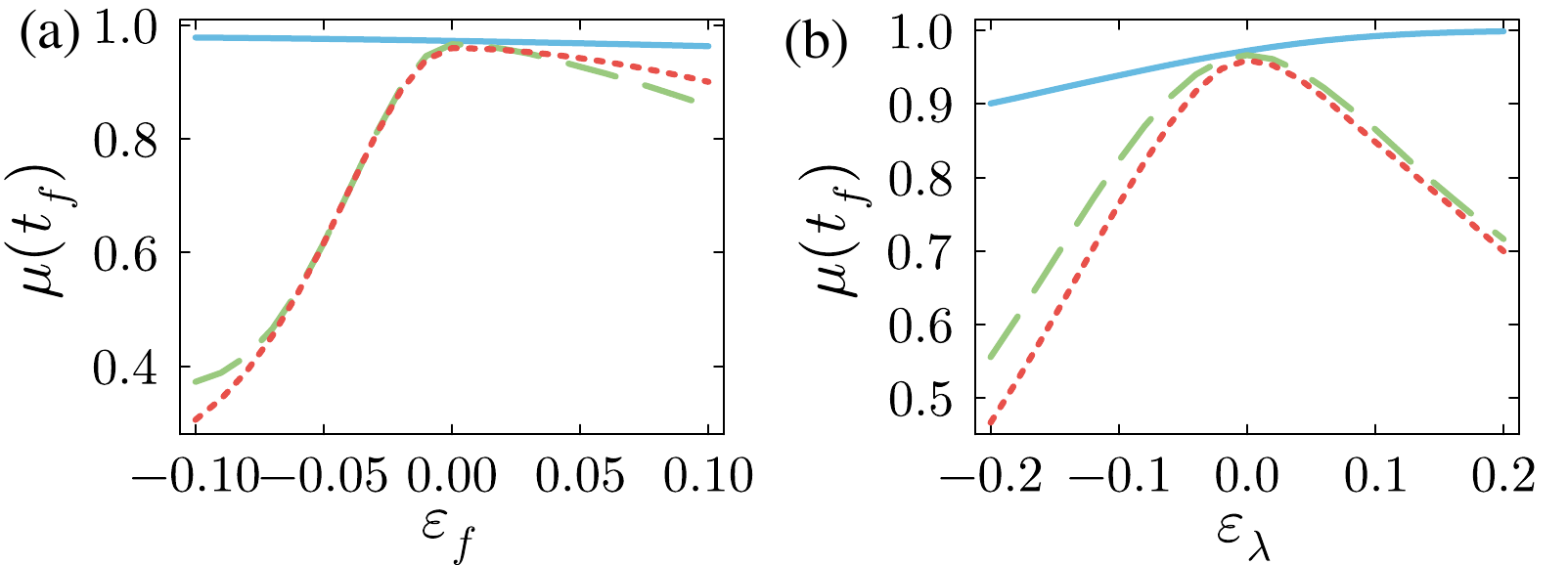}
\caption{
Sensitivity to static calibration errors. The normalized final magnetization \(\mu(t_f)\) is shown for the adiabatic reference (blue solid), GRAPE-optimized
protocol (green dashed), and optimized variational circuit
(red dotted) as a function of
(a) the Zeeman amplitude error \(\epsilon_f\) and
(b) the SOC amplitude error \(\epsilon_\lambda\).
In each scan, one control amplitude is rescaled while the other is kept
at its calibrated value, and the control profiles are not reoptimized.}
\label{fig:calibration_errors}
\end{figure}

\textit{Robustness to calibration errors.--}
Finally, we assess the sensitivity to static calibration errors by rescaling the
Zeeman and SOC controls as
$f_k\rightarrow(1+\epsilon_f)f_k$ and
$\lambda_k\rightarrow(1+\epsilon_\lambda)\lambda_k$, respectively.
Here, $\epsilon_f$ and $\epsilon_\lambda$ denote the dimensionless
fractional errors in the corresponding control amplitudes.
The errors are held fixed throughout the evolution and varied
independently, without reoptimizing the control profiles.
We scan $\epsilon_f\in[-0.1,0.1]$ and
$\epsilon_\lambda\in[-0.2,0.2]$; the resulting final normalized
magnetization $\mu(t_f)$ is shown in
Fig.~\ref{fig:calibration_errors}.

\jl{The long-duration adiabatic protocol is comparatively robust to calibration errors, whereas the shorter optimized protocols exhibit greater sensitivity, consistent with their reliance on more precisely shaped control amplitudes to achieve accelerated transfer.}
For Zeeman errors, at $\epsilon_f=-0.1$,
$\mu(t_f)$ decreases from 
\jl{its value at zero calibration error} to $0.373$ for GRAPE and
$0.306$ for the variational circuit. For SOC errors, the degradation is
weaker: at $\epsilon_\lambda=-0.2$, the corresponding values are
$0.555$ and $0.466$, respectively. The reference adiabatic evolution
shows a weaker dependence on calibration errors over the ranges
considered, consistent with its substantially longer evolution time.
Overall, the optimized protocols retain the SOC-assisted transfer under
moderate calibration errors but exhibit increased sensitivity,
particularly to negative Zeeman-amplitude errors.

\textit{Conclusion and outlook.--}
We have established a control principle for symmetry-constrained
many-body dynamics in the half-filled FH model.
Without SOC, conservation of total magnetization partitions the Hilbert
space into dynamically disconnected sectors, so that changes of the
instantaneous ground-state magnetization occur through
symmetry-protected crossings that cannot be followed dynamically.
A transient SOC control removes this obstruction by coupling neighboring
magnetization sectors and converting the crossings into avoided
crossings with finite gaps, thereby opening a pathway for
magnetization-sector transfer. Once this pathway is available, GRAPE
optimization of the Zeeman and SOC controls strongly accelerates the
transfer in both one- and two-dimensional FH systems, while the same
mechanism survives fermion-to-qubit mapping, Trotterization, and direct
variational circuit optimization. Thus, SOC determines dynamical
accessibility, whereas optimal control determines the timescale.

The required ingredients are compatible with existing experimental
platforms. \jl{Both the FH model and synthetic SOC have already
been experimentally realized in ultracold-fermion and optical-lattice
platforms}~\cite{Cheuk2012SOC,Kolkowitz_2017,Bromley2018SOC,
PhysRevLett.129.090403}. FH Hamiltonians have also been
engineered in gate-defined semiconductor quantum-dot arrays
~\cite{Hensgens_2017}, where electrically tunable SOC-induced spin-flip
tunneling provides a complementary control resource
~\cite{fernandez2024flying}. Programmable fermionic simulators and
digital quantum processors further provide routes toward the
circuit-level implementation considered here
~\cite{3nx4-bnyy,alam2025programmabledigitalquantumsimulation}.  More broadly, this framework points to transient symmetry breaking as a
general strategy for reshaping dynamical accessibility in constrained
many-body systems, with optimal control setting the timescale of the
resulting state transfer.

\textit{Acknowledgments.--}
This work is partially supported by PID2021-126273NB-I00, PID2023-
149072NBI00, and PID2024-157842OA-I00, Projects in the field of Artificial Intelligence 2025 (AIA2025-163435-C44) and the Severo Ochoa Centres of Excellence program through Grant CEX2024-001445-S.
R.Q.X. and J.L.T. appreciate the support from China Scholarship Council (CSC) under Grant Nos: 202206890001 and 202306890004. 
G.P. also acknowledges support
from CSIC Research Platform PTI-001 and
National Project No. QTP2021-03-002

\bibliography{bibfile}

\section*{----End Matter----}

\textit{GRAPE implementation.--}
The GRAPE gradient is evaluated using forward-propagated states
\begin{equation}
|\psi_k\rangle
=
\hat U_k\cdots\hat U_1|\psi_0\rangle ,
\end{equation}
and backward-propagated costates,
\begin{equation}
|\chi_k\rangle
=
\hat U_{k+1}^{\dagger}\cdots
\hat U_{N_T}^{\dagger}
\hat{\mathcal M}
|\psi(t_f)\rangle .
\end{equation}
For a control amplitude
\(v_{a,k}\in\{f_k,\lambda_k\}\), the gradient of the
final-magnetization objective in Eq.~\eqref{eq:grape_objective} is
\begin{equation}
\frac{\partial\mathcal L}{\partial v_{a,k}}
=
2\,{\rm Re}
\left[
\left\langle
\chi_k
\left|
\frac{\partial\hat U_k}{\partial v_{a,k}}
\right|
\psi_{k-1}
\right\rangle
\right].
\label{eq:endmatter_grape_gradient}
\end{equation}
For sufficiently small \(\delta t\), we approximate
\begin{equation}
\frac{\partial\hat U_k}{\partial v_{a,k}}
\simeq
-i\delta t\,\hat H_a\hat U_k,
\qquad
\hat H_a\in
\{\hat H_Z,\hat H_{\rm SOC}\}.
\label{eq:endmatter_grape_derivative}
\end{equation}
The gradients are supplied to a bounded L-BFGS-B optimizer.
Control bounds, initialization, and convergence criteria are given
in~\cite{SM}.

\textit{Variational circuit implementation.--}
At the \(k\)th Trotter layer, the fixed FH evolution is combined with
variational Zeeman and SOC controls. The corresponding circuit layer is
written as
\begin{equation}
\mathcal U_k(\boldsymbol{\theta}_k)
=
\mathcal U_{\rm SOC}(\theta_{k,{\rm SOC}})
\mathcal U_Z(\theta_{k,Z})
\mathcal U_{\rm C}(U\delta t)
\mathcal U_{\rm h}(\tau\delta t),
\label{eq:endmatter_trotter_layer}
\end{equation}
where
\(
\boldsymbol{\theta}_k
=
(\theta_{k,Z},\theta_{k,{\rm SOC}})
\)
denotes the variational Zeeman and SOC pulse areas at layer $k$.
Each block \(\mathcal U_\alpha\) is obtained after the
Jordan--Wigner transformation and decomposition of the corresponding
Hamiltonian term into elementary Pauli rotations.

The full circuit is
\begin{equation}
\mathcal U(\boldsymbol{\theta})
=
\mathcal U_{N_T}(\boldsymbol{\theta}_{N_T})
\cdots
\mathcal U_2(\boldsymbol{\theta}_2)
\mathcal U_1(\boldsymbol{\theta}_1).
\label{eq:endmatter_total_circuit}
\end{equation}
The circuit parameters are optimized using the same final-magnetization objective as in the Hamiltonian-level optimization,
\begin{equation}
\mathcal L_{\rm circ}(\boldsymbol{\theta})
=
\langle\psi_0|
\mathcal U^\dagger(\boldsymbol{\theta})
\hat{\mathcal M}
\mathcal U(\boldsymbol{\theta})
|\psi_0\rangle .
\label{eq:endmatter_circuit_objective}
\end{equation}
Further details of the gate decomposition and circuit optimization are given in ~\cite{SM}.

\end{document}


\crefname{equation}{Eq.}{Eqs.}
\crefname{figure}{Fig.}{Fig.}
\crefname{appendix}{Appendix}{Appendix}
\renewcommand{\thefigure}{S\arabic{figure}}
\renewcommand{\theequation}{S\arabic{equation}}
\renewcommand{\bibnumfmt}[1]{[S#1]}
\renewcommand{\citenumfont}[1]{S#1}
\renewcommand{\thesection}{S\arabic{section}}
\newcommand{\jl}[1]{{\color{black} {#1}}}

\title{
Supplementary Materials for: \\ Spin-Orbit-Mediated Magnetization Transfer in the Fermi-Hubbard Model
}

\author{Jialiang Tang$^{\orcidlink{0009-0001-6420-6492}}$}
\affiliation{Quantum Advanced Research Center (QuARC), CSIC, 28049, Madrid, Spain}
\affiliation{Instituto de Ciencia de Materiales de Madrid (ICMM), CSIC, 28049, Madrid, Spain}
\affiliation{\mbox{Departamento de Física Teórica de la Materia Condensada, Universidad Autónoma de Madrid, 28049 Madrid, Spain}}

\author{Ruoqian Xu$^{\orcidlink{0009-0005-3242-9674}}$}
\affiliation{Quantum Advanced Research Center (QuARC), CSIC, 28049, Madrid, Spain}
\affiliation{Instituto de Ciencia de Materiales de Madrid (ICMM), CSIC, 28049, Madrid, Spain}
\affiliation{\mbox{Departamento de Física Teórica de la Materia Condensada, Universidad Autónoma de Madrid, 28049 Madrid, Spain}}

\author{Gloria Platero$^{\orcidlink{0000-0001-8610-0675}}$}
\affiliation{Quantum Advanced Research Center (QuARC), CSIC, 28049, Madrid, Spain}
\affiliation{Instituto de Ciencia de Materiales de Madrid (ICMM), CSIC, 28049, Madrid, Spain}

\author{Yue Ban$^{\orcidlink{0000-0003-1764-4470}}$}
\email{yue.ban@csic.es}
\affiliation{Quantum Advanced Research Center (QuARC), CSIC, 28049, Madrid, Spain}
\affiliation{Instituto de Ciencia de Materiales de Madrid (ICMM), CSIC, 28049, Madrid, Spain}

\author{Xi Chen$^{\orcidlink{0000-0003-4221-4288}}$}
\email{xi.chen@csic.es}
\affiliation{Quantum Advanced Research Center (QuARC), CSIC, 28049, Madrid, Spain}
\affiliation{Instituto de Ciencia de Materiales de Madrid (ICMM), CSIC, 28049, Madrid, Spain}
\date{\today}
    
\maketitle
\tableofcontents

\section{Ground-state preparation}
\label{app:ground_state_preparation}

This section provides details of the initial-state preparation used in
the main text. We consider two complementary strategies. The
first is an analog adiabatic-loading protocol, which prepares the
antiferromagnetically correlated half-filled Fermi--Hubbard (FH) ground state
from a staggered N\'eel product state. The second is a CD-inspired
variational circuit that approximates the same target ground state for
the digital simulations. In both cases, the prepared state lies in the
zero-magnetization sector and serves as the initial state for the
subsequent SOC-assisted magnetization-sector transfer.

\subsection{Analog adiabatic loading protocol}
\label{app:analog_loading}

We consider analog adiabatic loading for both a six-site open chain
and a $2\times3$ lattice with open boundary conditions.
The time-dependent loading Hamiltonian is
\begin{equation}
    \hat{H}_{\mathrm{load}}(t)
    = [1-s(t)]\hat{H}_{\mathrm{stag}} + s(t)\hat{H}_{\mathrm{FH}},
    \qquad s(t)=t/t_f,
\end{equation}
with
\begin{align}
    \hat{H}_{\mathrm{stag}}
    &= -\chi\sum_i \eta_i
    \left(n_{i\uparrow}-n_{i\downarrow}\right), \\
    \hat{H}_{\mathrm{FH}}
    &= -\tau\sum_{\langle i,j\rangle,\sigma}
\left(c_{i\sigma}^{\dagger}c_{j\sigma}
    +\mathrm{h.c.}\right)
    +U\sum_i n_{i\uparrow}n_{i\downarrow}.
\end{align}
Here, $\eta_i=\pm1$ alternates between the two sublattices and
$\langle i,j\rangle$ denotes nearest-neighbor bonds.
We fix $\chi/\tau=2$ and set $\hbar=1$.

The initial state is the N\'eel product state selected by
$\hat{H}_{\mathrm{stag}}$. During the loading process, the staggered field is
gradually removed while the hopping and interaction terms are switched on,
such that $\hat{H}_{\mathrm{load}}(t_f)=\hat{H}_{\mathrm{FH}}$.
Since both spin populations are conserved during the loading protocol,
the evolution remains in the sector
$N_\uparrow=N_\downarrow=3$, corresponding to half filling and
zero total magnetization.

The preparation accuracy is quantified by the ground-state fidelity
\begin{equation}
    \mathcal{F}
    = \left|\langle\psi_{\mathrm{gs}}
    \mid\psi(t_f)\rangle\right|^2,
    \label{eq:supp_ground_state_fidelity}
\end{equation}
where $|\psi_{\mathrm{gs}}\rangle$ denotes the normalized ground state
of $\hat{H}_{\mathrm{FH}}$ in the corresponding particle-number sector,
obtained using DMRG, and $|\psi(t_f)\rangle$ is the normalized final
state after the loading protocol.
The real-time dynamics is simulated using two-site TDVP with
$\tau\Delta t=0.05$, a maximum bond dimension of $400$, and a
truncation cutoff of $10^{-10}$.
Although the present system sizes are also accessible to exact
diagonalization, we use DMRG and TDVP for state preparation to retain
a methodology applicable beyond exact-diagonalization sizes, while
exact diagonalization is employed below for the spectral analysis.

Figure~\ref{fig:ground_state_fidelity}(a,b) shows the resulting infidelity
$1-\mathcal{F}$ for both geometries.
For both the six-site chain and the $2\times3$ lattice, the infidelity
decreases with increasing ramp duration, consistent with the suppression
of diabatic excitations in the adiabatic limit.
As the interaction strength increases, the loading becomes progressively
slower, and a longer evolution time is required to achieve the same target
fidelity.
This behavior indicates that the characteristic adiabatic timescale grows
with $U/\tau$ in both geometries.

\subsection{CD-inspired variational preparation}

We next describe the digital preparation of the interacting FH ground
state. After the Jordan--Wigner transformation, the fermionic operators
are mapped onto qubit operators as
\begin{align}
c_j^\dagger
&=
\frac{1}{2}
\left(
\prod_{m=0}^{j-1} Z_m
\right)
\left(
X_j - iY_j
\right), \\
c_j
&=
\frac{1}{2}
\left(
\prod_{m=0}^{j-1} Z_m
\right)
\left(
X_j + iY_j
\right),
\end{align}
where $X_j$, $Y_j$, and $Z_j$ are Pauli operators acting on the $j$-th qubit. The Jordan--Wigner string $\prod_{m=0}^{j-1} Z_m$ enforces the correct fermionic anticommutation relations, and the resulting FH Hamiltonian is expressed
as a sum of Pauli strings suitable for digital implementation.

For the digital preparation, we consider an interpolation from the
noninteracting hopping Hamiltonian to the target interacting FH
Hamiltonian,
\begin{equation}
\hat H(s(t))
=
(1-s(t))\hat H_h
+
s(t) \hat H_{\mathrm{FH}},
\label{eq:adiabatic_path_fh}
\end{equation}
\jl{where \(\hat H_h\) is the noninteracting hopping Hamiltonian without the
on-site Coulomb interaction, and \(\hat H_{\rm FH}\) is the target
half-filled interacting FH Hamiltonian.}
The interpolation parameter satisfies $s(0)=0$ and $s(t_f)=1$. The initial state is chosen as the ground state of $\hat H_h$, which can be efficiently prepared using a sequence of Givens rotation gates~\cite{tang2024exploring}. The objective is to transform this state into the ground state of the interacting Hamiltonian $\hat H_{\mathrm{FH}}$.

To reduce the circuit depth associated with direct adiabatic
preparation, we construct a counterdiabatic (CD)-inspired variational
ansatz. The CD Hamiltonian is determined by the adiabatic
gauge potential, which suppresses diabatic transitions between
instantaneous eigenstates. We approximate the gauge potential using the
nested-commutator (NC) expansion~\cite{kolodrubetz2017geometry,sels2017minimizing,
claeys2019floquet},
\begin{equation}
\label{eq:cd}
\hat{\mathcal{A}}_s^{(\ell)}
=
i
\sum_{k=1}^{\ell}
\alpha_k(s)
\underbrace{
\left[
\hat H(s),
\left[
\hat H(s),
\cdots
\left[
\hat H(s),
\partial_s \hat H(s)
\right]
\cdots
\right]
\right]
}_{2k-1\ \mathrm{nested\ commutators}},
\end{equation}
where $\alpha_k(s)$ are variational coefficients and $\ell$ is the truncation order. In the limit $\ell\rightarrow\infty$, the expansion
recovers the exact adiabatic gauge potential
~\cite{claeys2019floquet}.

\begin{figure}[t]
\centering
\includegraphics[width=15cm]
{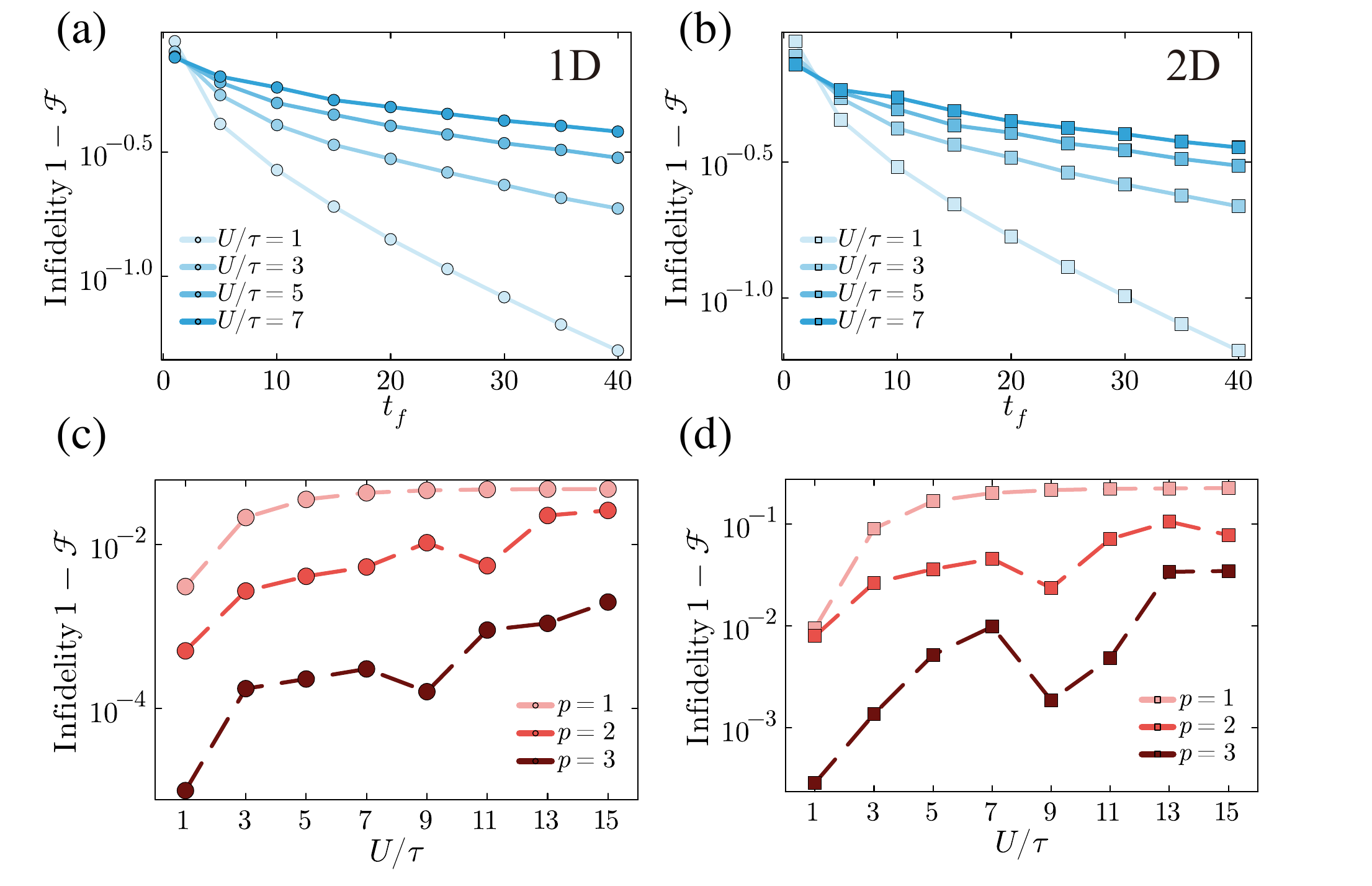}
\caption{
Ground-state preparation for the six-site FH chain and
the $2\times3$ lattice.
(a),(b) Ground-state infidelity $1-\mathcal F$ obtained from analog
adiabatic loading as a function of the ramp duration for different
interaction strengths $U/\tau$.
(c),(d) Ground-state infidelity relative to the target FH ground state
for the CD-inspired variational circuit as a function of the circuit
depth $p$.
Panels (a),(c) correspond to the six-site chain, while
(b),(d) correspond to the $2\times3$ lattice.
}
\label{fig:ground_state_fidelity}
\end{figure}

Rather than implementing the continuous CD Hamiltonian
directly, we use the first-order NC expansion to define a compact
variational operator pool. Specifically, we decompose
\begin{equation}
i[\hat H(s),\partial_s\hat H(s)]
=
\sum_i c_i(s)\hat{\mathcal A}_i^{(1)},
\end{equation}
where $\{\hat{\mathcal A}_i^{(1)}\}$ denote the independent operator
structures generated by the first-order commutator. These operators are
then used as generators of the variational circuit,
\begin{equation}
\hat U_{\rm CD}(\boldsymbol{\theta})
=
\prod_{r=1}^{p}
\prod_i
\exp\left(
-i\theta_{r,i}\hat{\mathcal A}_i^{(1)}
\right),
\label{eq:cd_ansatz}
\end{equation}
where $p$ is the circuit depth and
$\boldsymbol{\theta}=\{\theta_{r,i}\}$ are layer-dependent variational
parameters. Thus, the gauge-potential construction is used to determine
the operator content of the ansatz, while the corresponding coefficients
are optimized independently.
The variational loss function is defined as the expectation value of the
target FH Hamiltonian,
\begin{equation}
E(\boldsymbol{\theta})
=
\bra{\psi_0}
\hat U_{\mathrm{CD}}^\dagger(\boldsymbol{\theta})
\hat H_{\mathrm{FH}}
\hat U_{\mathrm{CD}}(\boldsymbol{\theta})
\ket{\psi_0},
\label{eq:cd_vqa_energy}
\end{equation}
where $\ket{\psi_0}$ is the ground state of the hopping Hamiltonian
$\hat H_h$. The optimized state is benchmarked against the target FH ground state
using the fidelity defined in Eq.~\eqref{eq:supp_ground_state_fidelity}.

As shown in Fig.~\ref{fig:ground_state_fidelity}(c,d), increasing the
circuit depth systematically improves the ground-state preparation in
both geometries. Over the interaction range considered, the infidelity
decreases as the number of variational layers is increased from
$p=1$ to $p=3$, with reductions of up to several orders of magnitude.
At $p=3$, the first-order NC-inspired operator pool provides a
high-fidelity approximation to the interacting ground state in both the
six-site chain and the $2\times3$ lattice. These optimized states are
therefore used as the digital approximations to the correlated FH ground
states that initialize the SOC-assisted magnetization-sector-transfer
protocols in the main text.

\begin{figure}[b]
\centering
\includegraphics[width=15cm]{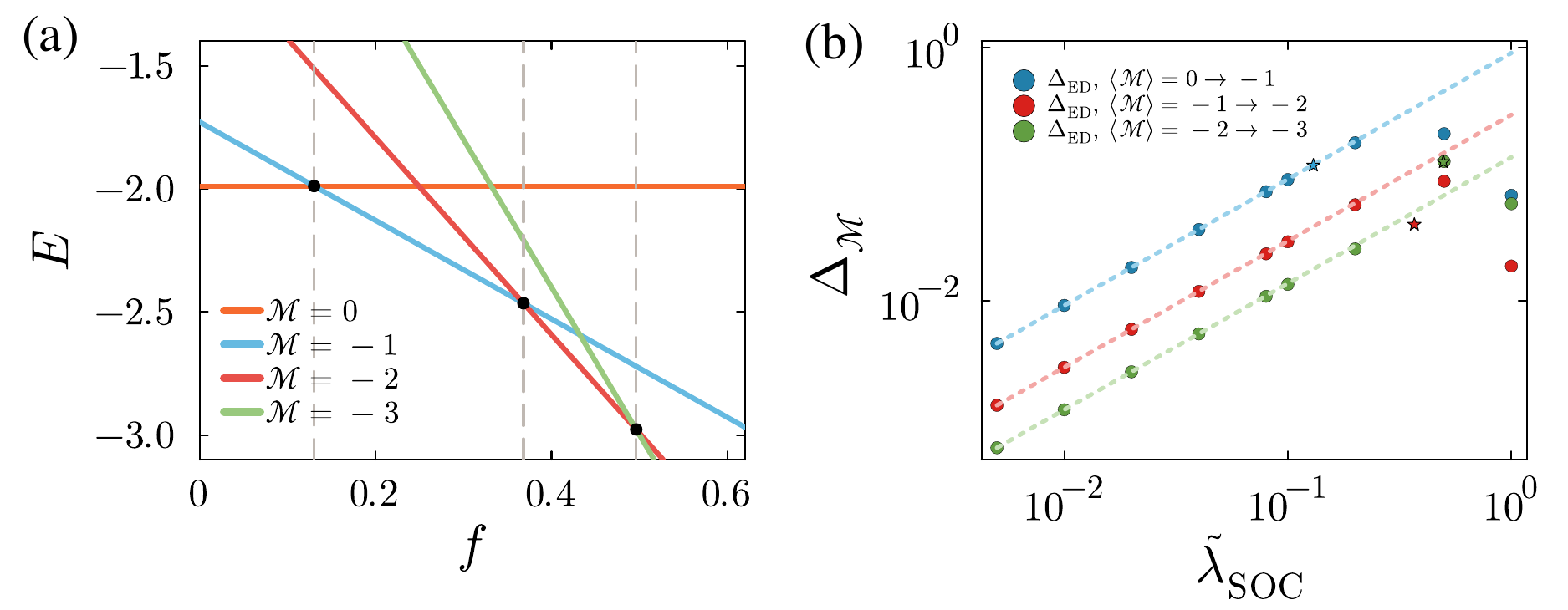}
\caption{
Two-level verification of SOC-induced sector transfer.
(a) Bare energies of the six-site half-filled FH chain without SOC, resolved by magnetization sector. The Zeeman ramp produces exact crossings along Eq.~\eqref{eq:phase_transition}.
(b) Log--log plot of the avoided-crossing gaps
$\Delta_{\mathcal M}$ obtained from full ED as functions of
$\tilde{\lambda}_{\rm SOC}$ \jl{at $t=t_c$}. Symbols show the full-ED results, while dotted
lines show the two-level prediction
in Eq.~\eqref{eq:prediction_1d}. Star-shaped markers indicate the
parameter values corresponding to the level transitions.
}
\label{fig:soc_two_level_verification}
\end{figure}

\section{Effective two-level description of SOC-induced sector transfer}

\subsection{One-dimensional chain}

To clarify the mechanism behind the SOC-assisted magnetization transfer, we first consider the Hamiltonian in the absence of SOC,
\begin{equation}
\hat H_0(t)=\hat H_{\rm FH}+f(t)\hat H_Z,
\qquad
\hat H_Z=2B\hat{\mathcal M}.
\end{equation}
Both the FH Hamiltonian and the Zeeman term conserve the total magnetization,
\begin{equation}
[\hat H_{\rm FH},\hat{\mathcal M}]
=
[\hat H_Z,\hat{\mathcal M}]
=0.
\end{equation}
The instantaneous eigenstates can therefore be classified by their magnetization quantum number \(\mathcal M\). The Zeeman ramp shifts the energy of each sector by an amount proportional to \(\mathcal M\), but it does not mix different sectors. As a result, crossings between states belonging to different magnetization sectors remain exact in the absence of SOC.

This structure provides a simple way to understand the role of SOC.
Consider a crossing between two bare eigenstates in adjacent
magnetization sectors, denoted by \(|\alpha,\mathcal M\rangle\) and
\(|\beta,\mathcal M-1\rangle\). Here, \(\mathcal M\) denotes the
magnetization quantum number, while \(\alpha\) and \(\beta\) label
eigenstate branches within the corresponding fixed-\(\mathcal M\)
sectors. The SOC-free
instantaneous energies are
\begin{equation}
E_{\alpha,\mathcal M}(t)
=
E_{\alpha,\mathcal M}^{(0)}
+
2Bf(t)\mathcal M,
\qquad
E_{\beta,\mathcal M-1}(t)
=
E_{\beta,\mathcal M-1}^{(0)}
+
2Bf(t)(\mathcal M-1),
\end{equation}
where \(E_{\alpha,\mathcal M}^{(0)}\) and
\(E_{\beta,\mathcal M-1}^{(0)}\) are the corresponding eigenenergies of
\(\hat H_{\rm FH}\) in the absence of both the Zeeman ramp and SOC.
The bare crossing time \(t_c\) is determined by
\(E_{\alpha,\mathcal M}(t_c)
=
E_{\beta,\mathcal M-1}(t_c)\), or equivalently
\begin{equation}
2Bf(t_c)
=
E_{\beta,\mathcal M-1}^{(0)}
-
E_{\alpha,\mathcal M}^{(0)} .
\label{eq:bare_crossing_condition}
\end{equation}

The SOC Hamiltonian contains spin-flip hopping terms and therefore
couples magnetization sectors differing by one unit. The relevant matrix
element between the two bare states is
\begin{equation}
g_{\mathcal M}
=
\langle
\beta,\mathcal M-1
|
\hat H_{\rm SOC}
|
\alpha,\mathcal M
\rangle .
\end{equation}
The gap opens provided that \(g_{\mathcal M}\neq0\). Near the crossing, the dynamics can be projected onto the two-dimensional
subspace spanned by these two bare states. In the basis
\(\{|\alpha,\mathcal M\rangle,|\beta,\mathcal M-1\rangle\}\), the
effective Hamiltonian is
\begin{equation}
\hat H_{\rm eff}(t)
=
\begin{pmatrix}
E_{\alpha,\mathcal M}(t) & \tilde{\lambda}_{\rm SOC}\times g_{\mathcal M}^{*} \\
\tilde{\lambda}_{\rm SOC}\times g_{\mathcal M} & E_{\beta,\mathcal M-1}(t)
\end{pmatrix}.
\end{equation}
where $\tilde{\lambda}_{\rm SOC}=\lambda(t)\lambda_{\rm SOC}$  denotes the effective time-dependent SOC envelope used
in the full ramp dynamics. Therefore, at the bare crossing
time \(t_c\), the avoided-crossing gap is
\begin{equation}
\Delta_{\mathcal M}(t_c)
\simeq
2\tilde{\lambda}_{\rm SOC} \jl{(t_c)}
\left|
g_{\mathcal M}
\right|.
\label{eq:prediction_1d}
\end{equation}
\jl{Here, the weak-SOC regime refers to the case where the SOC-induced
coupling to states outside the nearly degenerate two-level subspace
remains small compared with their energy separation from this subspace.
Under this condition, those additional states contribute only perturbative
corrections. }
Thus, in the weak-SOC regime, the induced gap is linear in the 
effective SOC strength \(\tilde{\lambda}_{\rm SOC}\), with a slope determined by the matrix element \(g_{\mathcal M}\).

We verify this picture numerically for the six-site half-filled system.
In the absence of SOC, the lowest-energy branches cross successively as
\begin{equation}
\mathcal M=0\rightarrow -1\rightarrow -2\rightarrow -3 .
\label{eq:phase_transition}
\end{equation}
As shown in Fig.~\ref{fig:soc_two_level_verification}(a), for \(B=1\), the crossings occur at
\(f_c=0.1301,\,0.3681,\,0.4962\), with corresponding SOC couplings
\(2|g_0|=0.9125\), \(2|g_{-1}|=0.2953\), and
\(2|g_{-2}|=0.1362\).
The weakest coupling is therefore associated with the final
\(\mathcal M=-2\rightarrow-3\) transition, which constitutes the main
bottleneck of the transfer.
To test the two-level prediction directly, we compute the local avoided gap by full exact diagonalization of the Hamiltonian including SOC at each bare crossing time $t_c$. As illustrated in Fig.~\ref{fig:soc_two_level_verification}(b), for sufficiently weak SOC, the extracted gaps follow
$\Delta_{\rm ED}\simeq2\tilde{\lambda}_{\rm SOC}|g_{\mathcal M}|$, confirming the
two-level prediction. Deviations become visible as the SOC strength
increases, reflecting multilevel hybridization beyond the perturbative
two-level regime.

Finally, near each crossing, the diabatic detuning is approximately
linear in time, so that the local dynamics admits a Landau--Zener
description, an approach that has also proved useful for many-body
avoided crossings in finite Fermi--Hubbard systems~\cite{triadu2025probing}.
Table~\ref{tab:two_level_crossing_parameters} summarizes the parameters
of the three consecutive avoided crossings for
$\lambda_{\rm SOC}=1$ and a total ramp time $t_f=1200$.
For each transition, $t_c$ denotes the position of the local
minimum gap, and
$\Delta_{\mathcal M}$ is the corresponding avoided-crossing
gap. The diabatic slope is defined as
\[
v_{\mathcal M}
=
\left|
\frac{d}{dt}
\left[
E_{\mathcal M}^{(0)}(t)
-
E_{\mathcal M-1}^{(0)}(t)
\right]
\right|_{t=t_c},
\]
where $t_c$ denotes the corresponding crossing position in the
absence of SOC.
We use $\Delta_{\mathcal M}^2/v_{\mathcal M}$ as a local
adiabaticity measure. Its value decreases from $115.502$ for the
$\mathcal M=0\rightarrow-1$ crossing to $20.1546$ for
$\mathcal M=-1\rightarrow-2$ and $4.5086$ for
$\mathcal M=-2\rightarrow-3$.
The final crossing therefore has by far the smallest local
adiabaticity measure and constitutes the dominant dynamical
bottleneck.

\begin{table*}[t]
\centering
\caption{
Avoided-crossing parameters obtained from the full instantaneous
many-body spectrum for $\lambda_{\rm SOC}=1$ and $t_f=1200$.
The diabatic slopes $v_{\mathcal M}$ are evaluated from the corresponding
SOC-free sector energies.
}
\label{tab:two_level_crossing_parameters}
\begin{tabular}{ccccc}
\hline\hline
Crossing
& $t_c$
& $\Delta_{\mathcal M}$
& $v_{\mathcal M}$
& $\Delta^2_{\mathcal M}/v_{\mathcal M}$ \\
\hline
$\mathcal M=0\rightarrow-1$
& $386.54$
& $0.52052$
& $2.3458\times10^{-3}$
& $115.5020$ \\
$\mathcal M=-1\rightarrow-2$
& $534.75$
& $0.28068$
& $3.9089\times10^{-3}$
& $20.1546$ \\
$\mathcal M=-2\rightarrow-3$
& $598.17$
& $0.13616$
& $4.1122\times10^{-3}$
& $4.5086$ \\
\hline\hline
\end{tabular}
\end{table*}

\begin{figure}[h]
\centering
\includegraphics[width=14cm]{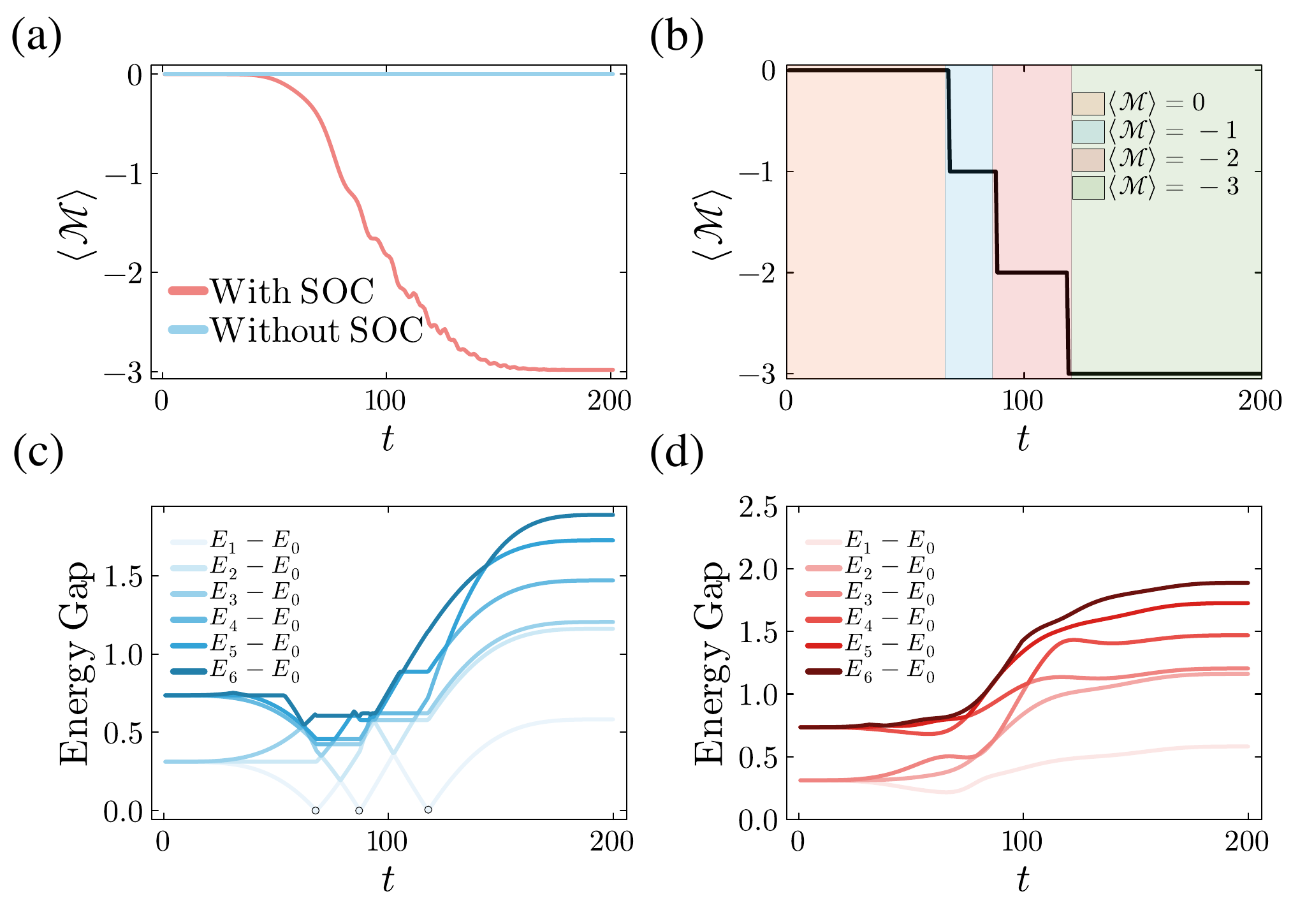}
\caption{
SOC-assisted magnetization-sector transfer in the two-dimensional
FH lattice.
(a) Time evolution of the magnetization
\(\langle \hat{\mathcal M} \rangle\) along the adiabatic path for the
\(2\times3\) lattice, with and without SOC.
(b) Magnetization quantum number \(\mathcal M\) of the instantaneous
ground state of the two-dimensional Hamiltonian
\(\hat H^{2D}(t)\) for \(\lambda_x(t)=\lambda_y(t)=0\), obtained by exact
diagonalization.
(c),(d) \jl{Instantaneous low-energy excitation spectrum,
\(E_n(t)-E_0(t)\) for \(n=1,\ldots,6\), along the adiabatic path
without and with SOC, respectively.} The gap closing in the absence of
SOC reflects symmetry-protected crossings between different
magnetization sectors, whereas SOC converts these crossings into avoided
crossings and opens a finite-gap pathway.
Parameters: \(U=7\) and \(t_f=200\) for the reference adiabatic
evolution.
}
\label{fig:2d_energy_gap}
\end{figure}

\subsection{Two-dimensional lattice}
The same sector-transfer mechanism is observed in the two-dimensional
FH lattice, as shown in Fig.~\ref{fig:2d_energy_gap} for a
half-filled \(2\times3\) system. Without SOC, the evolving state remains
confined to the initial \(\mathcal M=0\) sector, so that
\(\langle\hat{\mathcal M}\rangle=0\) throughout the ramp
[Fig.~\ref{fig:2d_energy_gap}(a)]. Nevertheless, the instantaneous ground state of the two-dimensional
Hamiltonian follows the same sequence of magnetization-sector changes
given in Eq.~\eqref{eq:phase_transition},
as shown in Fig.~\ref{fig:2d_energy_gap}(b). Since the FH and Zeeman
terms conserve \(\hat{\mathcal M}\), these changes occur through exact
level crossings when
\(\lambda_x(t)=\lambda_y(t)=0\), leading to gap closings along the
adiabatic path [Fig.~\ref{fig:2d_energy_gap}(c)]. In contrast,
transient SOC terms along the two lattice directions hybridize states
belonging to neighboring magnetization sectors, convert the exact
crossings into avoided crossings, and open finite gaps
[Fig.~\ref{fig:2d_energy_gap}(d)].

In these simulations we use the same units as in the main text,
setting \(\hbar=\tau_x=\tau_y=1\). Thus, \(U=7\),
\(\lambda_{\rm SOC}=1\), and \(t_f=200\) correspond to
\(U=7\tau\), \(\lambda_{\rm SOC}=\tau\), and
\(t_f=200\hbar/\tau\), respectively, where
\(\tau_x=\tau_y\equiv\tau\). For a representative cold-atom tunneling
scale \(\tau/h\sim160~{\rm Hz}\), these values correspond to
\(U/h\sim1.1~{\rm kHz}\), \(\lambda_{\rm SOC}/h\sim160~{\rm Hz}\), and
\(t_f\sim0.2~{\rm s}\). Compared with the 1D chain, the \(2\times3\) lattice exhibits a larger
minimum SOC-induced gap in our finite-size simulations, consistent with
the shorter reference time required to reach the target magnetization.

To make this finite-gap mechanism more explicit, we again project the
dynamics near each SOC-induced avoided crossing onto the local two-level
subspace associated with two neighboring magnetization sectors. Near a crossing between neighboring sectors, the relevant SOC matrix
element is
\begin{equation}
g_{\mathcal M}^{2D}
=
\left\langle
\psi_{\mathcal M-1}
\left|
\hat H_{\rm SOC}^{x}
+
\hat H_{\rm SOC}^{y}
\right|
\psi_{\mathcal M}
\right\rangle ,
\end{equation}
where \(|\psi_{\mathcal M}\rangle\) denotes the corresponding bare
eigenstate in the absence of SOC. For equal SOC envelopes
\(\lambda_x(t)=\lambda_y(t)=\lambda(t)\), projecting onto the
two nearly degenerate states gives the weak-SOC avoided-crossing gap
\begin{equation}
\Delta_{\mathcal M}^{2D} (t_c)
\simeq
2\tilde{\lambda}_{\rm SOC} (t_c) |g_{\mathcal M}^{2D}|.
\end{equation}
Thus, as in the one-dimensional chain, the SOC-induced gaps provide a
connected adiabatic pathway from the antiferromagnetically correlated
sector toward the fully polarized sector. This confirms that the
SOC-catalyzed sector-transfer mechanism is not restricted to a
one-dimensional geometry, but also applies to the two-dimensional
FH lattice.

\section{GRAPE optimal-control protocols}

This section summarizes the GRAPE optimal-control calculations. We first
introduce the full-control extension, where all Hamiltonian amplitudes
are optimized independently, and then provide the numerical details of
the GRAPE implementation used for both the restricted- and full-control
protocols.

\subsection{Full-control extension}
\label{app:full_control_grape}

In the main text, we considered a restricted-control GRAPE protocol in
which the FH parameters are held fixed and only the Zeeman
and SOC controls are optimized. This setting isolates the acceleration
that can be obtained by shaping the same control resources used in the
SOC-assisted adiabatic path. Here we consider a more flexible full-control setting, motivated by
recent progress in programmable FH simulators, where hopping
amplitudes, interaction strengths, and external potentials can be tuned
dynamically~\cite{3nx4-bnyy}. Accordingly, all Hamiltonian coefficients
entering the SOC-assisted protocol are promoted to independent
piecewise-constant controls.

The full-control Hamiltonian in the \(k\)-th time slice is
\begin{equation}
\hat H_k^{\rm full}
=
\tau_k \hat H_{\rm h}
+
u_k \hat H_{\rm C}
+
f_k \hat H_Z
+
\lambda_k \hat H_{\rm SOC},
\label{eq:full_control_hamiltonian}
\end{equation}
where \(\tau_k\), \(u_k\), \(f_k\), and \(\lambda_k\) denote the
layer-dependent hopping, interaction, Zeeman, and SOC amplitudes,
respectively. The corresponding time evolution is implemented using a first-order
Trotter decomposition,
\begin{equation}
\hat U_{\rm full}(t_f)
=
\prod_{k=1}^{N_T}
\left[
e^{-i\delta t\,\lambda_k\hat H_{\rm SOC}}
e^{-i\delta t\,f_k\hat H_Z}
e^{-i\delta t\,u_k\hat H_{\rm C}}
e^{-i\delta t\,\tau_k\hat H_{\rm h}}
\right],
\qquad
t_f=N_T\delta t .
\label{eq:full_control_evolution}
\end{equation}
The optimization objective is the same final-magnetization objective
used in the restricted-control protocol,
\begin{equation}
\mathcal L_{\rm full}(\boldsymbol{\theta})
=
\langle \psi(t_f)|
\hat{\mathcal M}
|\psi(t_f)\rangle ,
\qquad
|\psi(t_f)\rangle
=
\hat U_{\rm full}(t_f)|\psi_0\rangle ,
\label{eq:full_control_objective}
\end{equation}
with
\begin{equation}
\boldsymbol{\theta}
=
\left\{
\theta_{\tau,k},
\theta_{u,k},
\theta_{f,k},
\theta_{\lambda,k}
\right\}_{k=1}^{N_T},
\label{eq:full_control_parameter_vector}
\end{equation}
where the optimized variables are the integrated pulse areas
\begin{equation}
\theta_{\tau,k}=\delta t\,\tau_k,\qquad
\theta_{u,k}=\delta t\,u_k,\qquad
\theta_{f,k}=\delta t\,f_k,\qquad
\theta_{\lambda,k}=\delta t\,\lambda_k .
\label{eq:pulse_area_definition}
\end{equation}
Here, \(\tau_k\), \(u_k\), \(f_k\), and \(\lambda_k\) are the corresponding
piecewise-constant Hamiltonian amplitudes. The control profiles shown in
the figures are reported as amplitudes, obtained from the optimized pulse
areas by dividing by \(\delta t\).
For the negatively polarized target sector considered in this work,
minimizing $\mathcal L_{\rm full}$ drives the system toward the
ferromagnetic sector with $\mathcal M_{\rm tar}=-3$.

\begin{figure}[t]
 \centering
\includegraphics[width=12cm]{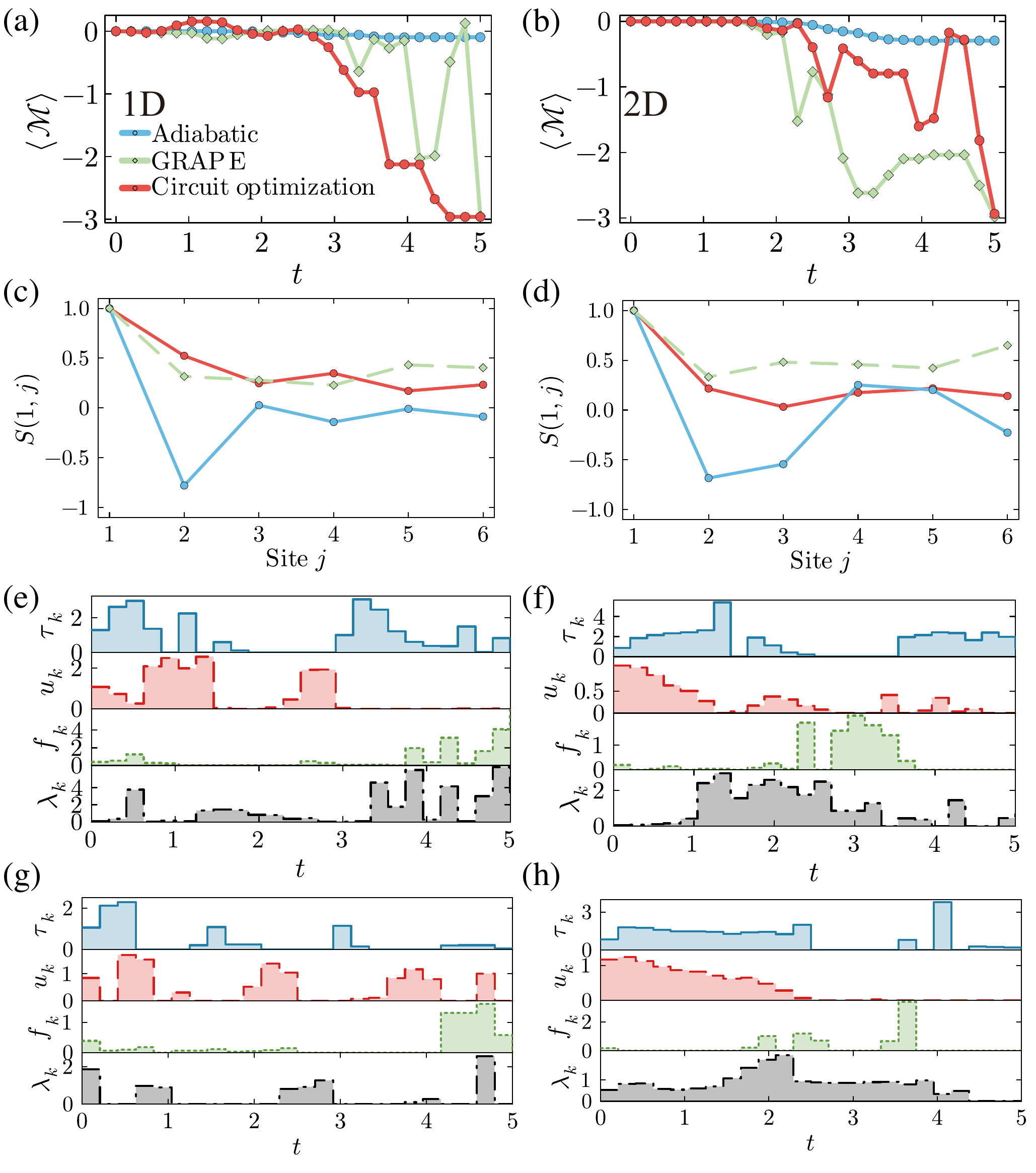}
\caption{
Optimized full-control profiles for SOC-assisted magnetic-state
conversion. (a),(b) Time evolution of the magnetization expectation value
$\langle \hat{\mathcal M} \rangle$ under the reference adiabatic path defined in Eq.~(3) of the main text, the GRAPE-optimized protocol, and the optimized
digital quantum circuit. (c),(d) Spin
correlations at the final time $t_f$, showing the change from
AFM-like correlations to FM-like alignment induced by the optimized
magnetic-field and SOC controls. (e),(f) Full-control GRAPE-optimized Hamiltonian amplitudes
\(\{\tau_k,u_k,f_k,\lambda_k\}\) for the 1D chain and the 2D
\(2\times3\) lattice, respectively, corresponding to the hopping,
on-site interaction, Zeeman-field, and SOC controls. 
(g),(h) Full-parameter variational-circuit-optimized 
Hamiltonian amplitudes
\(\{\tau_k,u_k,f_k,\lambda_k\}\) 
for the corresponding hopping, interaction, Zeeman, and SOC evolution
blocks in the 1D chain and the 2D \(2\times3\) lattice, respectively. All four
Hamiltonian terms are optimized independently over \(N_T=25\) time slices
with \(\delta t=0.2\), giving \(t_f=5\).
}
\label{fig:full_control}
\end{figure} 

Figure~\ref{fig:full_control} summarizes the performance of the
full-control protocol in both the 1D chain and the $2\times3$ lattice.
Fig.~\ref{fig:full_control}(a,b) shows that, within the short evolution time
$t_f=5$, the full-control GRAPE protocol drives the magnetization
rapidly toward the target polarized sector, achieving
$\mu(t_f) = 0.983$ in the 1D chain and
$\mu(t_f)= 0.994$ in the $2\times3$ lattice.
In contrast, the reference adiabatic ramp remains far from the
target on the same timescale. The optimized digital circuit
exhibits similar magnetization dynamics, demonstrating that rapid
sector transfer is also achievable with digital control.
Fig.~\ref{fig:full_control}(c,d) show the connected spin correlations at the
final time. These correlations become predominantly positive,
consistent with enhanced ferromagnetic spin correlations.

The optimized pulse profiles are shown in
Figs.~\ref{fig:full_control}(e,f). In contrast to the restricted-control
protocol discussed in the main text, where only \(f_k\) and
\(\lambda_k\) are optimized, the full-control protocol independently
optimizes the four amplitudes
\(\{\tau_k,u_k,f_k,\lambda_k\}\). Their nontrivial time dependence shows
that the optimized dynamics does not simply follow the smooth reference
adiabatic ramp. Instead, the hopping and interaction controls reshape
the correlated many-body background, while the Zeeman and SOC controls
drive and mediate the magnetization-sector transfer. Compared with the
restricted-control protocol, this additional freedom further compresses
the transfer time, reaching the target magnetization within \(t_f=5\).
Thus, when programmable control of the FH parameters is
available, optimizing all four amplitudes provides an additional route
to accelerating SOC-catalyzed magnetization-sector transfer. Because the full-control calculation is intended as an upper-bound
benchmark on the acceleration achievable with programmable FH
parameters, we do not use it as the conservative resource-matched
comparison in the main text.

\subsection{Numerical details of the GRAPE optimization}
\label{app:grape_details}

We summarize the numerical settings used for the GRAPE results shown in
Fig.~3 of the main text. For the fixed-$\tau,U$ calculations,  the Zeeman and SOC
controls were optimized independently in each time slice. The
optimization variables correspond to the time-integrated control
coefficients, with the associated amplitudes obtained by dividing by
$\delta t$. The evolution within each time slice was generated by the
piecewise-constant Hamiltonian $\hat H_k$, with the same time
discretization used throughout the optimization. We used
$(t_f,\delta t,N_T)=(30,0.2,150)$ for the six-site chain and
$(30,0.15,200)$ for the $2\times3$ lattice. For the restricted-control calculations, the hopping and interaction
amplitudes were fixed to their reference values, while only the Zeeman
and SOC pulse areas were optimized. The bounds were
\begin{equation}
\theta_{f,k},\theta_{\lambda,k}\in[0,\delta t],
\end{equation}
which is equivalent to the amplitude bounds
\begin{equation}
f_k,\lambda_k\in[0,1].
\end{equation}
For the auxiliary full-control calculations, all four pulse areas
\(\theta_{\tau,k}\), \(\theta_{u,k}\), \(\theta_{f,k}\), and
\(\theta_{\lambda,k}\) were optimized simultaneously. We imposed only
positivity constraints,
\begin{equation}
\theta_{\tau,k},\theta_{u,k},\theta_{f,k},\theta_{\lambda,k}\geq0,
\end{equation}
equivalently
\begin{equation}
\tau_k,u_k,f_k,\lambda_k\geq0,
\end{equation}
with no upper bounds unless otherwise stated.

The controls were initialized from the same smooth reference profiles
used in the adiabatic protocol. Specifically, the initial integrated
coefficients were obtained by multiplying the corresponding reference
amplitudes at each time slice by $\delta t$, followed by a random
perturbation of magnitude $10^{-4}$ generated with the fixed seed $7$.
All optimizations were performed
using the L-BFGS-B algorithm as implemented in
\texttt{scipy.optimize.minimize}~\cite{2020SciPy-NMeth}, together with analytic GRAPE
gradients. The optimization tolerance was set to $10^{-8}$, with a
maximum of $500$ iterations. For the fixed-$\tau,U$ calculations, the
optimization was terminated once
$\langle\hat{\mathcal M}\rangle<-2.90$ was reached. Both the six-site
chain and the $2\times3$ lattice satisfied this criterion before the
iteration limit. State propagation and gradient evaluation were carried
out in the fixed-particle-number Hilbert-space sector using sparse
matrix exponentiation; no finite-difference gradients were used.

\section{Variational-circuit details}

This section summarizes the circuit-level implementation used for the
SOC-assisted magnetization-sector-transfer protocols. We first describe
the full-parameter variational circuit, in which all Hamiltonian blocks
are optimized at the circuit level. We then provide the logical
gate-count estimates, the measurement procedures used to evaluate
magnetization, spin correlations, interaction energy, and hopping
observables, and the numerical optimization details for the
variational-circuit parameters.

\subsection{Full-parameter optimization}

In the full-parameter variational circuit, the hopping, interaction,
Zeeman, and SOC evolution blocks are optimized independently at each
Trotter layer. We use the same pulse-area convention $\boldsymbol{\theta}$ as in
Eqs.~\eqref{eq:full_control_parameter_vector} and
\eqref{eq:pulse_area_definition}: the optimized variables are the
integrated pulse areas
\(\theta_{\tau,k}\), \(\theta_{u,k}\), \(\theta_{f,k}\), and
\(\theta_{\lambda,k}\). These pulse areas enter the digital circuit directly as variational
rotation-angle parameters: within each Hamiltonian block, the optimized
quantity \(\theta_{\alpha,k}\) multiplies the fixed Pauli coefficient
\(h_{\alpha j}\) and determines the rotation angle of the corresponding
Pauli gate.

After the Jordan--Wigner transformation, each Hamiltonian block is
expanded in Pauli strings,
\begin{equation}
\hat H_\alpha
=
\sum_j h_{\alpha j}P_{\alpha j},
\qquad
\alpha\in\{{\rm h},{\rm C},Z,{\rm SOC}\}.
\end{equation}
For an integrated pulse area \(\theta_{\alpha,k}\), the corresponding
circuit block is implemented using the first-order product formula,
\begin{equation}
\mathcal U_\alpha(\theta_{\alpha,k})
=
\prod_j
e^{-i\theta_{\alpha,k} h_{\alpha j}P_{\alpha j}}
\approx
e^{-i\theta_{\alpha,k}\hat H_\alpha}.
\end{equation}
The \(k\)-th full-parameter circuit layer is therefore
\begin{equation}
\mathcal U_k^{\rm full}
=
\mathcal U_{\rm SOC}(\theta_{\lambda,k})
\mathcal U_Z(\theta_{f,k})
\mathcal U_{\rm C}(\theta_{u,k})
\mathcal U_{\rm h}(\theta_{\tau,k}) .
\label{eq:supp_full_parameter_layer}
\end{equation}
The full circuit is
\begin{equation}
\mathcal U_{\rm full}(\boldsymbol{\theta})
=
\prod_{k=N_T}^{1}
\mathcal U_k^{\rm full}.
\end{equation}

Each Pauli rotation is further decomposed into elementary gates.
Within each Hamiltonian block, the rotation angles are determined by the
same layer-dependent pulse area \(\theta_{\alpha,k}\) and the fixed Pauli
coefficients \(h_{\alpha j}\). Thus, four independent parameters are
optimized per layer, while the circuit structure remains fixed. The circuit parameters are optimized using the same final-time
magnetization objective as in the Hamiltonian-level GRAPE protocol,
\begin{equation}
\mathcal L_{\rm circ}^{\rm full}(\boldsymbol{\theta})
=
\langle\psi_0|
\mathcal U_{\rm full}^{\dagger}(\boldsymbol{\theta})
\hat{\mathcal M}
\mathcal U_{\rm full}(\boldsymbol{\theta})
|\psi_0\rangle .
\label{eq:supp_full_parameter_objective}
\end{equation}
For the negatively polarized target sector, minimizing
\(\mathcal L_{\rm circ}^{\rm full}\) drives the final state toward
\(\mathcal M_{\rm tar}=-3\).

Unlike the fixed-FH variational circuit, where the hopping and
interaction amplitudes are kept fixed at their reference values
\(\tau\) and \(U\),the full-parameter circuit optimizes all four pulse areas
\(\{\theta_{\tau,k},\theta_{u,k},\theta_{f,k},\theta_{\lambda,k}\}\)
on an equal footing. This additional freedom allows the circuit to
reshape the correlated many-body background while simultaneously using
the Zeeman and SOC blocks to mediate magnetization-sector transfer. As shown in Fig.~\ref{fig:full_control} (a,b), at
\(t_f=5\), the full-parameter circuit reaches
\(\mu(t_f)=0.987\) in the 1D chain and
\(\mu(t_f)=0.978\) in the \(2\times3\) lattice. These results show that
high final polarization can be achieved on a short timescale by jointly
optimizing the hopping, interaction, Zeeman, and SOC controls. The
corresponding optimized profiles are shown in
Figs.~\ref{fig:full_control} (g,h).

\subsection{Logical gate-count estimates}

The circuit depth quoted here refers to the Trotterized SOC-assisted
dynamical circuit and excludes the separate initial-state preparation
step. Each Trotter layer contains hopping, interaction, Zeeman, and SOC
evolution blocks. The logical circuit depth therefore scales linearly
with the number of Trotter layers \(N_T\).

After the Jordan--Wigner transformation, the fermionic evolution
operators are decomposed into the elementary gate set
\(\{R_X,R_Z,H,\mathrm{CNOT}\}\). The following estimates are logical
pre-transpilation resource counts, excluding hardware-specific routing,
parallel scheduling, and further gate cancellation. For the 1D chain,
one Trotter layer contains 402 elementary gates, including 172 CNOT
gates. Thus, for \(N_T=150\), the circuit contains 60300 elementary
gates and 25800 CNOT gates. For the \(2\times3\) lattice, one Trotter
layer contains 674 elementary gates, including 364 CNOT gates. Thus,
for \(N_T=200\), the circuit contains 134800 elementary gates and
72800 CNOT gates.

For the full-parameter optimizations, the circuit structure is
unchanged; only the layer-dependent pulse areas are varied. Therefore, for a fixed
\(N_T\), optimizing the additional hopping and interaction amplitudes
does not increase the logical gate count.

\subsection{Measurement strategy}
\label{app:measurement}

After the Jordan--Wigner transformation, all observables are represented
as Pauli operators. We use an interleaved spin-orbital ordering in which
qubits $(2i,2i+1)$ encode the spin-up and spin-down occupations of site
$i$. Diagonal observables, including magnetization, spin correlations,
and on-site interaction energy, are measured directly in the
computational basis, while hopping terms require a basis rotation prior
to measurement.

\subsubsection{Computational-basis observables}
\paragraph{Magnetization.}
The total magnetization defined in the main text,
\begin{equation}
    \hat{\mathcal M}
    =
    \frac{1}{2}\sum_i
    \left(
    n_{i\uparrow}-n_{i\downarrow}
    \right),
\end{equation}
is diagonal in the computational basis and can therefore be extracted
directly from spin-resolved occupation measurements. Using
$n_q=(I-Z_q)/2$, the local spin operator and total magnetization become
\begin{equation}
    \hat S_i^z
    =
    \frac{1}{4}
    \left(
    -Z_{2i}+Z_{2i+1}
    \right),
    \qquad
    \hat{\mathcal M}
    =
    \sum_i \hat S_i^z .
    \label{eq:supp_magnetization_pauli}
\end{equation}
Thus, no additional basis rotation is required. For the qubit pair
$(2i,2i+1)$, the local spin expectation value is
\begin{equation}
    \langle \hat S_i^z\rangle
    =
    \frac{1}{2}
    \left(
    P_{10}^{(i)}-P_{01}^{(i)}
    \right),
\end{equation}
where $P_{ab}^{(i)}$ denotes the marginal probability of measuring the
bit string $ab$ on the two spin orbitals associated with site $i$.
Summing over all lattice sites gives
\begin{equation}
    \langle\hat{\mathcal M}\rangle
    =
    \frac{1}{2}
    \sum_i
    \left(
    P_{10}^{(i)}-P_{01}^{(i)}
    \right).
    \label{eq:supp_magnetization_measurement}
\end{equation}
This quantity is the principal observable used to track the transfer
between different magnetization sectors throughout the dynamics.

\paragraph{Spin correlations.}
While the total magnetization characterizes the global polarization of
the system, the spin--spin correlations provide spatial information
about the magnetic order. For two distinct sites $i\neq j$, the
spin--spin operator is mapped to
\begin{equation}
    \hat S_i^z\hat S_j^z
    =
    \frac{1}{16}
    \left(
    Z_{2i}Z_{2j}
    -Z_{2i}Z_{2j+1}
    -Z_{2i+1}Z_{2j}
    +Z_{2i+1}Z_{2j+1}
    \right).
    \label{eq:supp_spin_correlation_pauli}
\end{equation}
Since this operator is also diagonal in the computational basis, its
expectation value can be reconstructed from joint bit-string
probabilities. For the qubit ordering
$(2i,2i+1,2j,2j+1)$,
\begin{equation}
    \langle \hat S_i^z\hat S_j^z\rangle
    =
    \frac{1}{4}
    \left(
    P_{1010}
    +P_{0101}
    -P_{1001}
    -P_{0110}
    \right).
    \label{eq:supp_spin_correlation_probability}
\end{equation}
The connected spin correlation reported in the main text is then
\begin{equation}
    S(i,j)
    =
    \langle\hat S_i^z\hat S_j^z\rangle
    -
    \langle\hat S_i^z\rangle
    \langle\hat S_j^z\rangle .
    \label{eq:supp_connected_spin}
\end{equation}
Subtracting the product of the local magnetizations isolates genuine
two-site correlations from the contribution associated with the net
polarization. For $i=j$, the corresponding local spin fluctuation is
\begin{equation}
    \left\langle
    \left(\hat S_i^z\right)^2
    \right\rangle
    =
    \frac{1}{4}
    \left(
    P_{10}^{(i)}+P_{01}^{(i)}
    \right).
\end{equation}

\paragraph{Interaction energy.}
The on-site interaction term is likewise diagonal in the computational
basis and is directly related to the double-occupancy probability. For
site $i$,
\begin{equation}
    n_{i\uparrow}n_{i\downarrow}
    =
    \frac{1}{4}
    \left(
    I-Z_{2i}-Z_{2i+1}
    +Z_{2i}Z_{2i+1}
    \right).
    \label{eq:supp_double_occupancy}
\end{equation}
Its expectation value is simply the probability of simultaneously
occupying the spin-up and spin-down orbitals,
\begin{equation}
    \langle n_{i\uparrow}n_{i\downarrow}\rangle
    =
    P_{11}^{(i)}.
\end{equation}
The total interaction energy can therefore be reconstructed without any
basis rotation as
\begin{equation}
    E_U
    =
    U\sum_i P_{11}^{(i)}.
\end{equation}
Together with the hopping contribution discussed below, this provides
the measurement of the FH energy used in the variational
ground-state preparation.

\subsubsection{Non-computational-basis observables}

\paragraph{Hopping energy.}
The hopping term is not diagonal in the computational basis and
therefore requires a basis rotation prior to measurement. Under the
Jordan--Wigner transformation, hopping between orbitals $p<q$ becomes
\begin{equation}
    -\tau
    \left(
    c_p^\dagger c_q+c_q^\dagger c_p
    \right)
    \longrightarrow
    -\frac{\tau}{2}
    \left(
    X_p Z_{p+1}\cdots Z_{q-1}X_q
    +
    Y_p Z_{p+1}\cdots Z_{q-1}Y_q
    \right).
    \label{eq:supp_hopping_general}
\end{equation}
For nearest-neighbor sites in the one-dimensional interleaved
spin-orbital ordering, $q=p+2$, giving
\begin{equation}
    Q_{\rm h}
    =
    -\frac{\tau}{2}
    \left(
    X_p Z_{p+1}X_{p+2}
    +
    Y_p Z_{p+1}Y_{p+2}
    \right).
    \label{eq:supp_hopping_1d}
\end{equation}
The endpoint operator can be diagonalized by the basis transformation
\begin{equation}
    M_1=
    \begin{pmatrix}
        1 & 0 & 0 & 0 \\
        0 & -1/\sqrt{2} & 1/\sqrt{2} & 0 \\
        0 & 1/\sqrt{2} & 1/\sqrt{2} & 0 \\
        0 & 0 & 0 & 1
    \end{pmatrix},
\end{equation}
acting on qubits $(p,p+2)$. After this rotation, the hopping expectation
value is reconstructed from computational-basis probabilities as
\begin{equation}
    \langle Q_{\rm h}\rangle
    =
    \tau
    \left(
    P_{001}
    -P_{011}
    -P_{100}
    +P_{110}
    \right),
    \label{eq:supp_hopping_probability}
\end{equation}
where the probabilities refer to the ordering $(p,p+1,p+2)$.
For longer Jordan--Wigner parity strings, including those associated
with some bonds of the two-dimensional lattice, the same procedure is
used with the corresponding intermediate $Z$ string retained.

\subsection{Variational-circuit optimization}

The variational-circuit parameters were optimized with the same
L-BFGS-B optimizer and initialization strategy used in the GRAPE
calculations. The objective was the final-time magnetization expectation
value,
\(\mathcal L_{\rm circ}=\langle\hat{\mathcal M}\rangle_{t_f}\), which was
minimized to drive the state toward the negatively polarized
ferromagnetic sector. For the fixed-\(\tau,U\) circuits, only the Zeeman
and SOC pulse areas, \(\theta_{f,k}\) and
\(\theta_{\lambda,k}\), were optimized, while the hopping and interaction
pulse areas were fixed by the reference FH parameters.

For the full-parameter circuits, all four pulse areas
\(\theta_{\tau,k}\), \(\theta_{u,k}\), \(\theta_{f,k}\), and
\(\theta_{\lambda,k}\) were optimized simultaneously. Since the circuit
layout is fixed, the optimization changes only the rotation angles of the
existing Trotter blocks. The amplitudes shown in the optimized profiles
are obtained by dividing the optimized pulse areas by \(\delta t\). The
L-BFGS-B tolerance was set to \(10^{-6}\), and the optimization was
stopped once \(\mathcal L_{\rm circ}<-2.90\) was reached or when the
optimizer satisfied its internal stopping criterion.

\bibliography{bibfile}